# How Disruptive is Financial Technology?

Douglas Cumming[†], Hisham Farag[‡], Santosh Koirala[§] and Danny McGowan[*]

**Abstract**

We study whether Financial Technology (Fintech) disrupts the banking sector by intensifying competition for scarce deposits funds and raising deposit rates. Using difference-in-difference estimation around the exogenous removal of marketplace platform investing restrictions by US states, we show the cost of deposits increase by approximately 5.9% within small financial institutions. However, these price changes are effective in preventing a drain of liquidity. Size and geographical diversification through branch networks can mitigate the effects of Fintech competition by sourcing deposits from less competitive markets. The findings highlight the unintended consequences of the growing Fintech sector on banks and offer policy insights for regulators and managers into the ongoing development and impact of technology on the banking sector.

19 January 2026

We are grateful for helpful comments and suggestions received from Maria Savona (the editor), two anonymous reviewers, Alin Andries, Piotr Danisewicz, Elena Loutskina, Huyen Nguyen, Enrico Onali, Tomasz Piskorski, Arisy Fariza Raz, Klaus Schaeck, Chendi Zhang, Mandy Zhang and seminar and conference participants at the Financial Management Association Annual Conference, Cardiff University, Dublin City University, Kathmandu University School of Management, Lund University, Swansea University, University of Exeter, University of Nottingham, and the University of Sydney. None of the authors have a conflict of interest or financial and personal relationships with other people or organizations that could inappropriately influence (bias) their work.
[†] Email: dcumming@stevens.edu. Stevens Institute of Technology.
[‡] Email: h.farag@bham.ac.uk. University of Birmingham.
[§] Email: s.koirala@bham.ac.uk . University of Birmingham.
[*] Email: danny.mcgowan@durham.ac.uk. Durham University.

# 1. Introduction

Innovation has long been viewed as a force of creative destruction that reshapes competitive dynamics within an industry (Schumpeter (1950)). Over the past decade, the rapid advancement of digital technologies has catalyzed a wave of innovation in the financial sector and given rise to a vibrant ecosystem of financial technology (Fintech) firms that pose a challenge to traditional incumbents. The vast amount of personal information in financial markets enables digital start-ups to leverage data analytics, platform-based business models, and algorithmic decision-making to deliver financial services in more cost-efficient and timely ways (Scott et al. (2017), Gupta et al. (2023)). Despite their rapid growth (Claessens et al. (2018), Thakor (2020)), we know relatively little about how Fintech innovations affect the strategic positioning of banks. The need for research here is acute because practitioners and policy makers have scarce resources with which to base actions, and unlike in other industries, the failure of financial institutions has widespread repercussions on employment, innovation and growth in non-financial sectors due to lending relationships (Matutes and Vives (1996), Boyd and De Nicolo (2005), Dell'Ariccia and Marquez (2006), Allen et al. (2021), Raz et al. (2022)).

Motivated by these concerns, we evaluate the disruptive effect of Fintech marketplace lenders on banks. Disruptive innovation theory predicts that marketplaces challenge banks by targeting underserved markets with simpler, cheaper, and more accessible lending services (Christensen (1997), Yu and Hang (2010)). Initially overlooked by incumbents, Fintechs improve rapidly, and capture lending market share from traditional banks. However, to originate credit, Fintechs require funding. We conjecture that marketplaces constitute a new source of competition for small banks in deposit markets because they offer depositors a relatively higher return, leading to reallocations of funding from deposit accounts to marketplaces. The entry of a marketplace lender therefore pushes banks to raise equilibrium deposit rates to prevent a drain of funding that would compromise their ability to originate credit (Li et al. (2019), Bollaert et al. (2021)). Furthermore, small financial institutions are likely to be more strongly affected due to their greater reliance on deposits to finance their activities.

Local US banking markets are an ideal setting in which to study Fintech's unintended consequences. Under the Securities Act of 1933 and the Securities Exchange Act of 1934, US state securities regulators have authority to determine whether marketplace lenders may solicit funds from their citizens and businesses headquartered in the state on a case-by-case basis. Regulators impose marketplace investing restrictions due to concerns that borrowers' loan applications may contain fraudulent information that poses a risk to investors. Obtaining regulatory approval to source funds from in-state investors requires that a marketplace meets the demands of a state regulator's 'merit review' process by demonstrating that its data safeguarding and verification measures protect investors from fraudulent claims in marketplace borrowers' credit applications (Chaffee and Rapp (2012)). The removal of marketplace investing restrictions is due to regulators' concerns about protecting investors from fraud and losses and are plausibly exogenous with respect to banks' deposit costs, and conditions within the banking industry more generally (Chaffee and Rapp (2012)). Additionally, the comprehensive classification of banks into large and small financial institutions by the Federal Deposit Insurance Corporation helps us neatly identify small banks for whom fintech competition is likely to be a competitive threat to deposit sourcing.

Our empirical analysis exploits the entry of marketplace lenders across states and time following the removal of entry barriers. Using difference-in-difference estimation applied to bank branch-level deposit rate data, we find robust evidence that allowing platforms to solicit funds within the state leads to a 5.9% increase in the cost of deposits. Within the universe of small banks, relatively larger institutions that are more reliant on wholesale funding, are affected to a lower degree. Similarly, the cost of deposits increases relatively more among banks that operate a limited number of branches, consistent with branch networks mitigating competition for funds by sourcing deposits from regions where marketplaces do not operate. More granular tests reveal that deposit rates increase across deposit products, but that the economic magnitude is largest for certificates of deposits and money market accounts. Further analyses reveals that the removal of Fintech barriers do not lead to contractions in the supply of bank deposits. Hence, while Fintech intensifies competition for funding, setting higher interest rates stems deposit outflows.[1]

Our research design exploits the panel structure of the branch-level data to ensure that changes in deposit costs and quantities are not driven by confounding forces. Specifically, we include bank-quarter-year fixed effects in the estimating equations. We thus identify Fintech's effects through comparisons between branches owned by the same bank at the same point in time. In essence, we compare how deposit costs evolve between a branch in a state that removes marketplace investing restrictions versus a branch in a state that does not, where the branches belong to the same bank. This approach purges all time varying forces at the bank level as well as macroeconomic fundamentals that have been found to influence deposit demand elsewhere in literature (Saunders and Schumacher (2000)).

A series of robustness tests rule out confounds. Diagnostic checks show no pre-emptive anticipatory trends in the cost of deposits prior to the removal of marketplace investing restrictions, the parallel trends identifying assumption holds, and the treated and control units are comparable along observable dimensions. Placebo tests indicate the cost of deposits does not simultaneously increase among large banks or small financial institutions in states contiguous to those that remove marketplace investing restrictions. This makes it unlikely the findings reflect confounding observable or unobservable omitted variables since banks in neighboring states operate in similar environments. In addition, shocks to bank soundness, monitoring by creditors (Danisewicz et al. (2018, 2021)), regulatory monitoring (Agarwal et al. (2014)), equity crowd funding, credit risk (McGowan and Nguyen (2023)), and changes to competition and market power within the banking industry (Focarelli and Panetta (2004), Berger et al. (2020), Duqi et al. (2021), McGowan et al. (2024)) do not drive the inferences. Further tests show the deregulation of crowdfunding restrictions do not confound the results, while methodological checks demonstrate that staggered treatments do not explain the findings (Callaway and Sant'Anna (2021)).

Our paper contributes to a rapidly evolving body of research on Fintech lenders and their consequences on bank lending. Several articles find that Fintech loans are a substitute for bank lending in consumer credit and mortgage markets (Cornaggia et al. (2018), Fuster et al. (2019), Tang (2019)), because they can originate loans more cheaply and process credit applications faster due to their algorithmic business model (Philippon (2015), Buchak et al. (2018)). Bartlett et al. (2022) document similar patterns of discrimination between Fintech and traditional lenders in mortgage markets. Allen et al. (2021) provide an extensive review of the Fintech lending literature, while Frame et al. (2019) survey Fintech's contribution to technological change and innovation in banking. A related stream of research documents the real effects of Fintech lending. Marin and Vona (2023) show Fintech lending spurs the reallocation of employment across sectors. Danisewicz and Elard (2023) illustrate the importance of Fintech credit to households and find that consumer bankruptcy rates increase in the absence of Fintech lending, while Jiang et al. (2025) show occupations with higher exposure to fintech experience a net decline in job postings and employment, though both complementary and substitutive effects emerge across different sectors. Whereas most extant research examines how Fintech affects bank lending, our paper studies its influence on deposit markets. We know of no other article on this topic. Our research is important because disrupting banks' deposit base has implications on financial institutions' ability to extend credit and support economic growth. Marketplaces' disruptive effects may also warrant regulatory scrutiny where they destabilize banks' operations and trigger changes in funding costs that are relevant from both macroprudential and monetary policy perspectives (Claessens et al. (2018), Thakor (2020)).[2]

---

[1] Our results are potentially externally valid as the business models of the marketplace platforms operating during the sample period resemble those in other countries.

A parallel literature studies the effects of innovations on the banking sector. Scott et al. (2017) estimate the long-run impact of the adoption of SWIFT, a network-based technological infrastructure and set of standards for worldwide interbank telecommunication, and find it increases bank profitability with larger effects among smaller financial institutions. Related work by DeYoung et al. (2007) shows that internet adoption is associated with improvements in bank profitability, mainly through increased revenues from deposit service charges. Using cross-country data, Beck et al. (2016) find that different measures of financial innovation, capturing both a broad concept and specific innovations, are associated with faster bank growth, but also higher bank fragility and worse bank performance during the financial crisis. Audretsch et al. (2020) assess how various policy incentive mechanisms, including those related to Fintech, influence innovative start-ups.

---

[2] Another set of papers evaluates how other types of Fintech innovation affect banks. Pal et al. (2021) illustrate the disruptive effects of mobile payment technologies. Hornuf (2021) find that banks tackle digital innovations by cooperating with Fintech firms and that they take ownership stakes in small Fintechs but build product-based collaborations with large Fintech competitors. Elliehausen and Hannon (2024) also find evidence of collaboration between Fintechs and banks due to interest rate ceilings. Bollaert et al. (2021) review how crowdfunding and initial coin offerings, as well as Fintech lenders, affect access to finance. Papers describing the factors contributing to the growth of Fintech and peer-to-peer lending include Claessens et al. (2018), Thakor (2020), Berg et al. (2022), Erel and Liebersohn (2022), Griffin et al. (2023) and Balyuk et al. (2025).

Our evidence matters for policymakers. Fintech innovations are transforming the way financial services are provided. This opens opportunities to consumers, such as cheaper credit and improved financial inclusion among borrowers excluded by traditional financial institutions. However, the transformation comes with potential risks to consumers and investors, and more broadly, to financial stability and integrity, which regulators seek to mitigate. Optimizing the benefits while minimizing potential risks to the financial system poses a challenge to regulators as the Fintech sector often lies outside their remit and the speed of Fintech innovation makes it difficult for regulators to respond in a timely manner (Ehrentraud (2020)). Our research highlights that marketplaces erode banks' deposit base which has implications for lending, stability and the transmission of monetary policy. While the extent of disruption remains low, the remarkable growth of marketplace platforms warrants attention from bank managers and policymakers alike in future.

The paper is structured as follows. Section 2 provides an overview of the data set. We provide details of the regulatory environment surrounding marketplace investments, and the legal background to the marketplace investing restrictions at the heart of our empirical methods in Section 3. We outline the identification strategy in Section 4, and present econometric results in Section 5. Section 6 deals with alternative explanations and robustness tests. Finally, we draw conclusions in Section 7.

## 2. Data Description

The econometric analysis relies on branch-level data from two sources. The FDIC Summary of Deposits database reports annual information on the geographical location of each bank branch throughout the US. This allows us to observe the quantity of deposits held by branch *b* belonging to bank *i* located in state *s* during year *t*.

Deposit cost information is taken from Ratewatch.com. This source provides weekly deposit and loan rates for each deposit and loan product that a branch offers. Ratewatch.com reports deposit rates based on the funding rate (FR) and annual percentage yield (APY). The measures provide strongly similar values as shown in Table 2. Using this information, we calculate the quarterly deposit rate (FR) and annualized percentage yield (APY) which measure the average quarterly deposit rate across all deposit products, for each branch over 2004Q1 to 2019Q4.[3] Using the granular product-level information, we also calculate the quarterly deposit rate paid on interest checking (IC), regular savings (SAV), money-market (MM) and 12-month certificates of deposits (CD) accounts to provide detailed insights into some of the most important deposit products banks offer.

Fintech competition is unlikely to have substantial implications on large banks' deposit base due to their scale, geographical reach, and access to wholesale funding. Rather, it is small financial institutions that are most likely to experience more intense deposit competition as depositors reallocate funds to fintech platforms. Our tests thus focus on small banks and we use the list of large banks published by the FDIC on quarterly basis to remove large banks from the sample.[4]

We merge additional data taken from several sources. We retrieve quarterly bank-level data from the Federal Financial Institutions Examination Council 031 Condition and Income Reports (call report) database. This provides information from 2004Q1 to 2019Q4 on several bank variables including bank size (total assets), return on assets (ROA), total liabilities, and deposit liabilities. To capture local business cycles and demand-side determinants of deposit costs, we use the state-level per capita income growth rate (Bureau of Economic Analysis), population growth rate (Bureau of Economic

---

[3] We aggregate the monthly data to the quarterly level because we merge in bank level data that is available at quarterly intervals. Our choice to begin the sample in 2004Q1 is motivated by the fact that Prosper and Lending Club were incorporated in 2005 and 2006, respectively. Setting the starting point at 2004Q1 therefore provides sufficient time to test the parallel trends assumption.

[4] The FDIC publishes a quarterly list of large banks, defined as those with assets over $300 million (https://www.federalreserve.gov/releases/lbr/). As a robustness test, we also run the baseline models removing the top 100 banks by asset size rather than using the FDIC's list. The unreported results are similar, albeit the effect sizes are smaller.

Analysis), unemployment rate (Bureau of Labor Statistics), and the number of business establishments per capita (County Business Patterns). Table 1 provides a definition of each variable in the data set. Table 2 reports summary statistics.

[Insert Table 1: Variable Description]     [Insert Table 2: Descriptive Statistics]

As we detail below, investing restrictions on Lending Club and Prosper were removed at different times by each state. We contacted both platforms and each state securities regulator to verify the date when investing restrictions were removed. Using this information, we construct the variable, Fintech index$_{st}$, which is a count of how many platform investing restrictions have been removed in state $s$ in quarter $t$.

## 3. Institutional Background

Lending Club and Prosper are the most prominent marketplace lenders in the US and operate similar business models. Prospective borrowers register with a platform and complete an online loan application. Using digital screening algorithms, the platforms assign each application a credit risk rating that determines whether the loan is listed on the marketplace for funding. During the application process platforms screen the borrower's credit history, outstanding debt, income, employment status, and other risk factors. Applicants' risk rating determines the interest rate a borrower pays.[5]

Investors do not make direct loans to borrowers, rather an issuing bank issues the loan to the borrower and then sells the loan to the platform.[6] The platform then issues a separate note to the investor with a return on the investment contingent on the borrower repaying the original loan (Chaffee and Rapp, 2012). Platforms do not take a stake in each loan, rather they charge service fees for originating each loan and on trading notes between investors in the secondary market.

Most borrower applications are unsecured consumer loans. These are primarily used to consolidate existing debts, although a substantial share of loans is used for home repairs and to finance personal or family purchases. While business loans are increasingly common, they remain a minority. The interest rate on marketplace loans ranges between 6.46% and 29% on Lending Club and 6.95% and 35.99% on Prosper. Loan amounts range between $1,000 and $40,000 and the term structure varies between 12 and 60 months.

---

[5] Before 2010 Prosper operated an auction for each loan whereby investors would submit bids (an interest rate) for each loan. The lowest bidders would win the auction and funds from those bidders were pooled to extend loans. From 2010 Prosper shifted to a model like that described above.
[6] Lending Club and Prosper have both used WebBank as the issuing bank.

### 3.1 State Marketplace Investing Law

The notes that are offered, sold, and purchased in the marketplace lending model constitute securities and are regulated by the Securities Act of 1933 and the Securities Exchange Act of 1934.[7] The Acts mandate that securities are registered either with a federal or state regulator. Section 18(b) of the Securities Act of 1933 stipulates that securities that may be listed and trade on a national market system (a registered exchange) are exempt from state-level registration and may be federally registered. As marketplaces' notes are not listed or traded on a national market system, the platforms must secure approval from state securities regulators to solicit funds from investors in each state (Cornaggia et al., 2018).

Many state securities regulators mandate security registrants meet the requirements of a 'merit review'.[8] This requires the state securities regulator find that, "the business of the issuer is not fraudulently conducted…that the plan of issuance and sale of the securities…would not defraud or deceive" (Chaffee and Rapp, 2012).[9] Information provided by borrowers in loan applications may be inaccurate, missing, or deliberately misleading. For example, they may misstate their income, current employment status, or employment history. Where a marketplace is unable to verify the information in borrowers' loan applications, the regulator rules it is unable to conclude that the business is not fraudulently conducted as required by state law. In these cases, the platform is denied the opportunity to register securities by the state regulator and is prohibited from soliciting funds from investors within the state. Marketplaces are only granted approval to solicit funds in a merit review state once the state securities regulator is convinced the platform has implemented procedures that ensure investors cannot be defrauded (Chaffee and Rapp, 2012).

---

[7] Section 2(a)1 of the Securities Act of 1933 and section 3(a)10 of the Securities Exchange Act of 1934 provide the definition of a 'security'. Both sections include within the definition of a security the terms 'investment contracts' and 'notes'. Marketplace loans fall under this umbrella.
[8] The states are Alabama, Arizona, Arkansas, Indiana, Iowa, Kansas, Kentucky, Maine, Maryland, Massachusetts, Michigan, Nebraska, North Carolina, North Dakota, Ohio, Oklahoma, Tennessee, Texas, Pennsylvania, Vermont, Virginia, and West Virginia.
[9] Ohio is a representative example of the law in merit review states (Chaffee and Rapp, 2012). See, Section 1701.09 of the Ohio Revised Code and Amendments for further details.

[Insert Table 3: Timing of Deregulation across States]

The remaining states permit marketplace lending without restrictions. This is because these states' securities law mirrors the Securities and Exchange Commission's approach to securities offerings which does not involve merit review but simply requires disclosure (GAO, 2011).[10] As these states historically followed this approach, after their establishment marketplaces were immediately granted approval to solicit investment funds. Among these states, seven authorize investing in notes but only for 'sophisticated' investors that meet suitability requirements. This is the case for all securities, including marketplace loans.[11] In most of these states, investing is limited to individuals with an income of at least $70,000 and a minimum net worth of $70,000. California imposes less stringent requirements, and only for individuals who invest more than 10% of their wealth in notes. The reasons states impose these restrictions are the financial health of marketplace investors.[12] Table 3 provides an overview of the dates when each state security regulator removed investing restrictions for Lending Club and Prosper.

**3.2 Regulatory Exogeneity**

Our review of the legal literature shows the state-level marketplace investing restrictions are due to regulators' concerns about protecting investors from fraud. The restrictions are unrelated to the cost of bank deposits and conditions within the banking industry more generally. Changes in investing restrictions are driven by a platform completing the merit review process, that is, convincing state securities regulators that their procedures accurately verify borrowers' application claims and ensure that investors are not exposed to fraud. Merit review is a key regulatory tool which allows states to reject offerings deemed overly risky or unfair to investors, even if disclosure rules are met. For digital platforms, merit review is a regulatory burden that requires proof of investor protections before public solicitation. The removal of Fintech investing restrictions in merit review states thus hinges on whether the platform can convincingly document that investors are protected from fraud in credit applications. In states that do not have a merit review process, the removal of marketplace investing restrictions is due to federal SEC regulations that are unrelated to the marketplace lending and banking industries. Lending Club and Prosper are therefore able to solicit funds from investors in these states as soon as the platform goes live.

---

[10] See the Government Accountability Office report *Person-to-Person Lending: New Regulatory Challenges Could Emerge as the Industry Grows*, supra note 5. http://www.gao.gov/new.items/d11613.pdf.
[11] The states with suitability requirements are California, Idaho, Kentucky, New Hampshire, Oregon, Washington, and Virginia.
[12] For example, the Kentucky Department of Financial Institutions noted that Lending Club's auditor's "going concern" letter mentioned its negative earnings. The department opined that investment in the site "constitutes a level of risk suitable only to Accredited Investors" (Chaffee and Rapp, 2012).

Deregulation is thus primarily due to marketplace lenders providing convincing evidence to regulators that their screening procedures adequately protect investors from fraud in borrowers' credit applications. The removal of marketplace investing restrictions is thus independent with respect to deposit markets and the banking sector more generally, making these events plausibly exogenous with respect to the outcomes we study.

## 4. Research Design

### 4.1 Empirical Model

To isolate causal inferences, we use difference-in-difference estimation that exploits time-varying changes in marketplace investing restrictions across US states. We compare the cross-time evolution of the dependent variable in branches in states that remove marketplace investing restrictions relative to branches in states where investing barriers remain in place. We estimate

$$y_{ibst} = \beta Fintech\ index_{st} + \delta X_{ibst} + \varphi_i + \varphi_{bst} + \varepsilon_{ibst}, \qquad (1)$$

where $y_{bist}$ is a dependent variable (e.g. deposit costs) for branch $i$ which belongs to bank $b$ in state $s$ during quarter-year $t$. $Fintech\ index_{st}$ is a count variable of marketplaces soliciting funds in a state there. Higher values indicate more intense Fintech competition as multiple marketplaces operate within a jurisdiction, thereby offering greater opportunities for investors. $Fintech\ index_{st}$ takes the value 0 if neither Lending Club nor Prosper have been granted permission to solicit funds, and 1 (2) if one (two) of the platforms have been granted permission to solicit funds within the state. $X_{ibst}$ is a vector of control variables; $\varphi_i$ and $\varphi_{bst}$ denote branch and bank × quarter × year fixed effects, respectively; $\varepsilon_{ibst}$ is the error term. Owing to the multilevel structure of the panel data, we follow Vig (2013) and two-way cluster standard errors by bank and quarter-year. Similarly, for specifications containing data aggregated to the state-quarter level, we apply two-way clustering at the state and quarter-year levels.

While the review of the legal literature suggests the removal of investing restrictions are exogenous with respect to our outcomes of interest, we conduct empirical diagnostic checks to verify this assumption. Online Appendix Table 1.A reports estimates of

$$d_{st} = \beta X_{st} + \varphi_t + \varepsilon_{st}, \qquad (2)$$

where $d_{st}$ is a dummy variable equal to 1 in quarter $t$ if state $s$ removes investing restrictions on either Lending Club or Prosper; $X_{st}$ is a vector containing state-level variables (population, the unemployment rate, the corporate tax rate, the mean deposit rate across all bank branches in the state, the mean Z-score of all banks operating in the state, and mean bank size (measured as the natural logarithm of assets)); $\varphi_t$ denote quarter-year fixed effects; $\varepsilon_{st}$ is the error term.[13]

Intuitively, estimates of $\beta$ will be statistically significant if a state characteristic predicts deregulation. In Online Appendix Table 1.A we find no significant associations between a the state variables and the removal of marketplace investing restrictions irrespective of whether we estimate equation (2) using a Cox Proportional Hazards model or a Weibull Hazard model. State size and macroeconomic conditions, measured using population and unemployment rates, are insignificant while delinquency rates on auto, credit card, mortgage, and student debt are unrelated to the timing of deregulation. Importantly, we find no links between the characteristics of the banking sector and the removal of investing barriers. The deposit rate, bank soundness, and bank size coefficient estimates are all insignificant. The removal of investing restrictions is therefore independent with respect to deposit market conditions, and features in the banking sector more generally, suggesting that estimates of $\beta$ in equation (2) are unlikely to be driven by simultaneity bias.

---

[13] Equation (2) does not contain state fixed effects as this would lead to a singular matrix issue when estimating a duration model.

Equation (1) exploits the panel structure of the data set to rule out confounds. Specifically, we include bank-quarter-year fixed effects, $\varphi_{bst}$, in the estimating equation. This eliminates unobservable confounds that may influence the cost or quantity of deposit holdings, both in the cross-section and time-varying bank specific and aggregate forces. The bank-quarter-year fixed effects also sharpen identification since estimates of $\beta$ are computed through comparisons between branches owned by the same bank at the same point in time. The average treatment effect is thus estimated using comparisons of $y_{bist}$ between branches that belong to the same bank but are located in different states during the same quarter that are exposed to different fintech competition intensities. To confound the inferences, an omitted variable must therefore systematically correlate with the removal of marketplace lending restrictions in a state and deposit costs within a bank branch. This appears unlikely.

### 4.2 Diagnostic Tests

Before reporting econometric results, we test the identifying assumption underlying difference-in-difference estimation: parallel trends. To more formally inspect whether parallel trends holds in the lead up to the removal of restrictions on the separate platforms, we estimate the equation.

$$y_{bt} = \beta_j Fintech\ index_{st-j} + \delta X_{bt} + \varphi_b + \varphi_t + \varepsilon_{bt}, \qquad (3)$$

where all variables are defined as in equation (1) except $Fintech\ index_{st-j}$ (where $j \in (1,2,3,4)$) are the first, second, third, and fourth lag of $Fintech\ index_{st}$, and $\varphi_b$ and $\varphi_t$ denote bank and quarter-year fixed effects, respectively. Intuitively, one would expect the coefficients $\beta_1$ to $\beta_4$ to be statistically insignificant if the parallel trends assumption holds because investing restrictions change in quarter $t$ but not beforehand. Put differently, during each pre-treatment period there should be no statistically significant changes in the cost of deposits between branches in states that do and do not subsequently remove marketplace investing restrictions.

[Insert Table 4: Identifying Assumptions Test]

The results of these tests are reported in Table 4. Throughout all columns of the table the coefficient estimates are economically close to 0 and statistically insignificant.

Together the graphical and econometric evidence suggests that the parallel trends assumption holds, and the conditions for drawing valid inferences from a difference-in-difference estimator are met. The inferences lend further support to the view that banks do not anticipate the removal of marketplace investing restrictions and pre-emptively change their deposit pricing behavior.

[Insert Table 5: Pre-treatment Characteristics]     [Insert Table 6: Lending Test]

Difference-in-difference estimates are more credible where the treatment and control groups resemble each other before treatment. In conjunction with parallel trends, this adds credibility to the implied counterfactual. Table 5 therefore presents the results of *t*-tests on the equality of several bank-level characteristics prior to the removal of marketplace investing restrictions. We find no significant differences between the groups in terms of size, capitalization, profitability, or their branch networks. Leverage, the variance of return on assets and equity, and bank soundness (measured using the Z-score) are also highly comparable.

For marketplace investing restrictions to be salient in determining bank deposit rate setting requires that marketplace lending responds to the removal of investing restrictions. In essence, removing investing restrictions should provoke an increase in marketplace lending because a marketplace has access to more funding that it can deploy in credit markets. Consistent with this conjecture, the estimates in Table 6 show that the removal of investing restrictions provokes a significant increase in marketplace lending using state-level data. A one-unit increase in the Fintech index provokes a roughly 32% increase in the amount of credit marketplaces originate within a state.[14] Intuitively, as barriers to investing in a state are eliminated, marketplaces can obtain more funding which they use to fund loans. Deregulation of investing restrictions thus has a complementary effect on marketplace lending.

## 5. Results

---

[14] The average treatment effect is computed as $100 \times (e^{0.2831} - 1) \approx 32.7\%$.

In this section, we first present evidence on how the cost of deposits responds to fintech competition and then report estimates indicating these responses are due to a contraction in deposit supply.

**5.1 Fintech Competition and Deposit Costs**

[Insert Table 7: Fintech Competition and Banks' Cost of Deposits]

Panel A of Table 7 presents estimates of equation (1) using deposit rate (FR) as the dependent variable. In column 1, the *Fintech index* coefficient estimate is 0.0559 and is statistically significant at the 1% level. Removing investing restrictions on 1 platform therefore increases the cost of deposits by 2.97%.[15] Allowing two marketplace platforms to solicit funds within the state raises deposit costs by 5.94%. Among the control variables, we estimate the cost of deposits is significantly negatively associated with the personal income growth rate, population, establishments per capita, and the unemployment rate. There are no bank-level control variables included in the regression because they are captured by the bank-quarter-year fixed effects.

[Insert Figure 1: Interaction Weighted Estimates]

Recent econometric advances highlight that the strict exogeneity assumption may fail under the two-way fixed effect design in cases where treatment is staggered across time because the composite error term can correlate with the treatment variable and group fixed effects. We therefore follow the Callaway and Sant'Anna (2021) approach to correct for any potential bias and check the robustness of our findings to using an interaction-weighted dynamic difference-in-difference. We first normalize the date of treatment to the quarter in which marketplace investing restrictions are removed to 0. Online Appendix Table 2.A reports the econometric results of the Callaway and Sant'Anna (2021) approach while Figure 1 illustrates the point estimates and 95% confidence intervals from this test. During the five quarters preceding deregulation the average treatment on the treated point estimates are economically close to zero and statistically insignificant. The insignificant dynamic pre-treatment coefficients further indicate that bank branches do not anticipate the removal of investing restrictions and adjust deposit pricing in advance.

---

[15] The dependent variable in Table 7 is the natural logarithm of the quarterly branch-level deposit rate. The point estimate therefore implies that one platform increases deposit costs by $(e^{0.0559} - 1) \times 100\% = 5.75\%$ in column 1.

However, following removal of the restrictions, deposit costs in affected states significantly increase relative to the counterfactual. Staggered treatments do not appear to drive baseline inferences.

Prior research shows small banks have a comparative advantage in sourcing deposits (Stein (2002), Berger and Udell (2004)). In contrast, large financial institutions tend to rely more on wholesale market funds and can leverage their geographically dispersed branch networks to obtain deposits in lower cost markets (Gilje et al. (2016)). These characteristics may mean that small financial institutions face greater competitive pressure following the entry of marketplace lenders because they are more reliant on deposits to fund their activities. We therefore test for heterogeneity in the effect of investing deregulation across bank size and branch network structure.

The estimates in column 2 include interactions between the fintech index and bank size in equation (1). We continue to find the fintech competition is significantly positively related to the cost of deposits. However, the extent of this increase is inversely related to bank size. The fintech index–bank size interaction coefficient is negative and significant at the 5% level, showing that relatively larger banks' deposit funding costs are less affected by fintech competition, compared to smaller financial institutions.

Banks may also experience differential changes in deposit competition after the removal of investing restrictions according to how many branches they operate. A larger branch network potentially allows a bank to avoid deposit competition from marketplaces by sourcing deposits from geographies where competition is less intense. For example, a multi-state bank could avoid deposit competition in state A (where investing restrictions are removed) by sourcing deposits from state B (where restrictions remain in place). To capture this effect, we interact the number of branches variable with the fintech index and include this variable in equation (1). The evidence in column 3 supports the conjecture. The interaction coefficient is negative and significant at the 1% level. Following the lifting of investing restrictions, deposit rates increase less among banks with more extensive branch networks.

Column 4 reports estimates from an equation that interacts the Fintech index with the total amount of marketplace lending in each state during the quarters after the removal of investing restrictions. The purpose of this test is to ensure that the effect of new entry by marketplace lending platforms is stronger when such lending is more active post-deregulation. The results show that the removal of marketplace lending restrictions provokes a significant increase in banks' deposit costs that is equivalent in magnitude to the baseline estimates. Moreover, the interaction term's coefficient estimate is positive and significant, indicating that deposit costs respond stronger in states where marketplaces originate more credit. Economically, the interaction coefficient indicates that for each marketplace investing restriction removed, the marginal effect of marketplace lending on deposit rates increases by 0.0010 percentage points. In essence, banks respond more strongly to marketplace lending competition as platforms are granted approval to solicit funds in their state. Fintech presence amplifies the competitive pressure of marketplace lending in deposit markets.[16]

Panel B in Table 7 replicates Panel A but uses deposit rate (APY) as the dependent variable in equation (1). The inferences remain strongly similar both in terms of the coefficients' magnitude and statistical significance.

[Insert Table 8: Deposit Product-level Effects and Deposit Supply]

To obtain more granular insights into which deposit categories are affected by fintech competition, we estimate equation (1) using the deposit rate on various products and present the results in Table 8. Across the table, the entry of a marketplace lender significantly increases the interest rate paid by 2.45% on checking accounts (column 1), 9.49% on money market accounts (column 2), 7.08% on savings accounts (column 3) and 9.25% on CDs (column 4). Columns 5-8 report similar inferences using annual percentage yields rather than funding rates. We also study how marketplace deregulation influences deposit spreads (the difference between marketplace rate and average bank-level deposit rates). Given the average spread in the sample is 5.82%, the estimates in column 9 show a statistically significant 1.90% (-0.0011/0.0582) reduction in spreads following entry by a marketplace. We also study the effects of fintech competition on banks' loan spreads, measured as the difference between the average loan rate and average funding rate.

---

[16] We are grateful to an anonymous reviewer for suggesting this test.

Column 10 shows this margin also significantly contracts once Fintechs enter the market implying a reduction in bank's net interest margins and the profitability of lending.

In column 11, we conduct a validation check by estimating the Fintech index coefficient while controlling for bank loan pricing. The effect is stable and quantitatively similar to the baseline results. Finally, in column 12, we examine the effect of Fintech competition on deposit growth. The average treatment effect is insignificant. This suggests that competition from marketplaces forces banks to raise deposit rates, and that these pricing responses are effective in preventing deposit flight to marketplaces.

Together, the findings are consistent with marketplaces acting as a substitute investment for depositors. When marketplaces enter a market, depositors have an option to reallocate their funds from bank deposits to marketplaces with potential to reduce the supply of deposits. This forces banks to set higher equilibrium deposit interest rates to prevent a drain of liquidity.

### 5.2 Heterogeneity Analysis and Long-run Effects

Do the removal of marketplace investing restrictions affects deposit markets differently depending on the level of banking competition? A new competitor for funding is likely to exert a larger marginal effect in concentrated markets. Panel A in Online Appendix Table 3.A reports estimates of a model including an interaction between the Fintech index and an inverse Herfindahl-Hirshman index (that is 1 − HHI) of bank deposit market shares in a county. Deposit rates are significantly higher in more competitive markets as banks compete intensely for funding. The effect of removing marketplace investing restrictions is smaller in these markets, which aligns with ex-ante intense competition between banks limiting further pricing adjustments.

Relatedly, marketplaces are likely to provoke larger pricing changes in markets where there is a limited supply of deposits such as rural areas. We therefore include an interaction between the Fintech index and a dummy variable that indicates whether a branch is located in a rural county. Consistent with this intuition, in Panel B of Online Appendix Table 3.A we find that removing investing restrictions spurs a relatively larger increase in rural areas' deposit rates.

Do these Fintech-induced deposit market changes affect bank profitability? In Online Appendix Table 4.A we find that banks' return on assets falls in the face of deregulation, but there is no significant change to return on equity. However, there is a significant, albeit modest, increase in leverage. The effect on profitability suggests that while marketplace lenders provoke an increase in bank deposit costs, the economic magnitude of this effect does not substantially raise banks' overall funding costs, but as shown by the increase in leverage, they force banks to secure other forms of debt to finance their operations.

## 6. Alternative Explanations and Robustness Tests

Within our data set there are 79 separate instances where marketplace investing restrictions are removed (for example, restrictions on investing through Lending Club are removed in Arizona in 2015Q2 counts as one instance). To bias our results, an omitted variable must systematically coincide with each of the 79 distinct removal episodes. This is much less likely compared to a setting with only one treatment event. Nevertheless, we conduct a series of sensitivity checks to rule out other plausible explanations for the results.

**6.1 Placebo Tests**

Falsification exercises provide a window into potential observable or unobservable confounding factors. Specifically, in samples where there are no changes to fintech competition, we should observe no effects on deposit costs. To conduct placebo tests, we restrict the sample to states that are contiguous to, but did not remove marketplace investing restrictions at the same time as, state A. For example, Maryland removed investing restrictions on Lending Club at 2016Q1, but Delaware, Pennsylvania, Virginia, and West Virginia did not.[17] We therefore include banks from the contiguous states, and randomly assign banks to placebo treatment status, and the rest to placebo control status and estimate,

---

[17] We impose a restriction that a contiguous state can only be included in the placebo sample providing it has not removed any Fintech investing restrictions within the previous three years.

$$y_{bist} = \beta Placebo_{st} + \delta X_{bist} + \varphi_{bst} + \varepsilon_{bist}, \qquad (4)$$

where all variables are defined as in equation (1), except $Placebo_{st}$ which is a dummy variable equal to 1 indicating placebo treatment status, 0 otherwise.

[Insert Figure 2: Simulation of Placebo Tests]

Online Appendix Table 5.A reports estimates of equation (4) using the contiguous state donor pool (column 1), and samples where the share of banks randomly assigned to placebo treatment status ranges between 30% and 75% (columns 2 to 5). Encouragingly, the placebo coefficient estimate is statistically insignificant and economically close to zero in throughout all columns in Panel A. This is also the case in Panel B of the table when we instead use the APY measure of deposit costs.

Figure 2 presents corroborating evidence using a Monte Carlo simulation exercise on equation (4). We use 1,000 replications to ensure that the placebo results are not driven by certain samples.[18] The placebo coefficient estimate of $\beta$ is normally distributed and closely centered on 0, which it should be if the placebo treatment has no significant effect. Moreover, we reject the null hypothesis in 44 cases, consistent with the 5% type-1 error rejection rate one should obtain under the 5% significance level.

If our baseline findings capture confounding forces, and potential secular trends in the cost of bank deposits, the placebo coefficient estimates should be similar in economic magnitude and statistical significance to the baseline results. This is not the case across both tests, despite the placebo samples containing banks that operate in geographical proximity and being observationally equivalent along several dimensions to treated banks. It is thus highly unlikely that our findings reflect omitted variable bias.

We conduct two further falsification experiments to affirm the results. First, large banks' deposits are unlikely to be substantially affected by competition from Fintech lenders due to their relative size differences. Online Appendix Table 6.A presents estimates from a placebo test in which the sample comprises only large banks (defined as the top 100 by total assets). In both specifications the treatment effect is insignificant. Second, we study whether deregulation episodes that are independent with respect to deposit markets influence deposit markets. Intuitively, these shocks should have no effect unless the treatment we study is driven by omitted variables. Online Appendix Table 7.A shows the state-level legalization of cannabis has no significant effect on deposit rates.

---

[18] To ensure the placebo results are not due to a specific sampling decision, we repeated the analysis in equation (4) 1,000 times by randomly assigning 50% of banks in qualifying contiguous states to placebo treatment status and the rest to placebo control status. Given the null hypothesis of 0 effect is correct, we should only reject the null if we make type-1 errors. This is what we find. At the 5% significance level, we reject the null 4.4% of the time, in line with the type-1 error rate.

## 6.2 Alternative Explanations

The market discipline literature predicts that debtholders monitor bank risk taking and price such effects into debt security prices (Calomiris, (1999), Danisewicz et al. (2018, 2021)). Martinez Peria and Schmukler (2001) show that depositors monitor banks' condition and respond to risky actions by demanding higher deposit interest rates as compensation. For this channel to confound the inferences, a bank characteristic must differentially influence the outcome variable depending on the Fintech index because the bank-quarter-year fixed effects capture the direct effect. To ensure the increase in the cost of deposits we attribute to the removal of investing restrictions does not reflect debtholders demanding risk premia in response to changes in bank soundness and profitability, we include an interaction between the Fintech index and banks' Z-score as an additional control variable in equation (1) to capture distance to default. Despite this change, the Fintech index coefficient estimate in column 1 of Table 9 is similar in economic and statistical magnitude to before.

[Insert Table 9: Bank Condition and Debtholder Monitoring]

We also test whether profitability and the variance of bank returns drive our inferences as debtholders may respond to changes in a bank's condition. The results in columns 2 to 5 demonstrate this is not the case. In column 6 we consider whether shocks to leverage influence our findings but find this not so. Theory and evidence show that non-depositors are especially important monitors because they possess more sophisticated monitoring technologies relative to depositors (Birchler, 2000; Danisewicz et al., 2018). We follow Danisewicz et al. (2018) and approximate non-depositor monitoring using non-deposit liabilities' costs. Column 7 reports the estimates which show our key inferences remain similar to the baseline specifications.

[Insert Table 10: Market Power and Competition]

Higher deposit costs could be driven by changes in market power and competition within the banking industry. For example, new bank entrants increase demand for deposits (McGowan et al. 2024) while shocks to concentration may influence banks' pricing decisions. As before, the presence of the bank-quarter-year fixed effects means that to contaminate the inferences, these forces must differentially influence deposit costs depending on the Fintech index. We therefore include interactions between the Fintech index, banks' deposit market share within the state, and the Herfindahl-Hirschman index of bank deposit market competition in equation (1). The Fintech index coefficient remains stable in columns 1 and 2 of Table 10 despite the changes. In the remainder of the table, we consider whether changes in the cost of deposits reflects the competitive effects of entry and exit. The key findings are robust to including controls for the opening and closing of bank branches (columns 3 and 4) and the entry and exit of banks (columns 5 and 6) within the state.

[Insert Table 11: Industry Dynamics and Survivorship Bias]

Relatedly, by bidding up the cost of deposit funds marketplace lenders may erode banks' net interest margins leading marginal banks to fail which reduces entry incentives. To ensure the results do not reflect industry dynamics or survivorship bias, we test the robustness of our results to removing observations of banks that fail or enter during the sample. This also ensures that the results are not due to the secular decline in the number of banks through time. The estimates in Table 11 show that removing these observations has no bearing on the findings.

**6.3 Alternative Financial Intermediation and Sensitivity Checks**

The Federal government and US states also removed restrictions on equity crowdfunding during the sample period. The timing of these law changes does not systematically correlate with the removal of marketplace investing restrictions. To ensure our findings are not driven by reforms of other types of Fintech law, we include a further interaction between the Fintech index and a dummy variable that equals 1 if a state has removed restrictions on equity crowdfunding. The estimates in column 1 of Online Appendix Table 8.A are robust. Column 2 demonstrates that the results remain despite removing observations from after 2011 to exclude variation during the time period when equity crowdfunding was deregulated.

Marketplaces allow investors to invest in new loans listed on the marketplace (the primary market) but may also buy and sell notes listed on a secondary market provided by the platform. In some states investors that do not have a Lending Club account can access the secondary market on Lending Club (but not the primary market) through the third-party brokerage platform FolioFn. However, relatively few investors pursue this option because the FolioFn platform is difficult to operate and the secondary market is illiquid (Harvey (2018)). There is no systematic correlation between the removal of marketplace investing restrictions and the states where FolioFn operates. To ensure the findings are not driven by the entry of FolioFn into new markets, we append equation (1) with an interaction variable between a dummy that equals 1 if FolioFn operates in state $s$ during quarter $t$, 0 otherwise, and the Fintech index. In column 3 the baseline findings remain robust. The interaction coefficient is statistically insignificant, consistent with the small volume of funds directed through the platform.

Changes in marketplace investing restrictions may correlate with other types of entrepreneurial finance. While venture capital (VC) funding is typically directed towards firms, and not to the borrowers that use marketplace platforms, we append equation (2) with an interaction between the Fintech index and the per capita quantity of VC funds in each state-year to ensure VC activity does not drive our inferences. Despite including these controls the Fintech index coefficient reported in column 4 remains similar to the baseline estimates.

Prior research shows corporate tax rates may influence a bank's deposit pricing strategy (Demirguc-Kunt and Huizinga, 1999, 2010). We find in column 5 that our findings remain robust to controlling for the top state marginal corporate tax rate.

Reputational concerns may lead marketplace lenders to avoid regions with systematically higher rates of borrower default to ensure investors do not suffer high losses and withdraw their funds. We therefore include interactions between the Fintech index and the rate of default (that is, the share of loans that are 90+ days in arrears) on auto loans, mortgages, and student debt in each state-year as further control variables in equation (1). The Fintech index coefficient is robust in column 6.

Lastly, in column 7, we report estimates of equation (1) using a sample that includes observations only from 2011Q1 onwards. This ensures that the findings are not due to either the financial crisis, or the temporary closure of Prosper and Lending Club during 2008 when the Securities Exchange Commission issued cease and desist orders that compelled the platforms to change their business models to conform to securities regulation. The findings endure. Online Appendix Table 8.A reports similar inferences using the APY dependent variable.

Evidence suggests Fintechs collaborate with banks to circumvent state interest rate ceilings on consumer credit which could, in turn, limit competition in deposit markets. Elliehausen and Hannon (2024) argue that FinTech-bank partnerships are less likely to occur in states with either high or no ceiling on personal loan interest rates. To examine whether and to what degree our result is sensitive to partnerships, we follow Elliehausen and Hannon (2024) and estimate equation (1) for two subsamples: states with either high-rate or no ceiling on personal loans (defined by Elliehausen and Hannon (2024) as Alabama, Georgia, Idaho, Illinois, Indiana, Kentucky, Louisiana, Mississippi, Missouri, New Mexico, Oklahoma, South Carolina, Tennessee, Texas, and Wisconsin), and all other states. Online Appendix Table 10.A presents the results of this test. The estimates show the removal of marketplace investing restrictions produces a significant increase in deposit rates irrespective of whether a state has an interest rate ceiling. This suggests that any potential collaboration between Fintech lenders and banks does not influence the main results.

## 7. Conclusions

Cycles of innovation have repeatedly disrupted and transformed the financial intermediation market. Recently, new digital technologies have allowed marketplace lending platforms to rapidly expand credit supply. This poses a challenge to banks as these platforms source funds that could otherwise be deployed as deposits. We show that in the US, following the removal of marketplace investing restrictions small banks experience significant increases in the cost of deposits to defend against deposit flight.

Our findings have important policymaking implications. The Fintech revolution has led regulators to question the risks and advantages of financial technologies to borrowers, particularly with respect to over indebtedness and bankruptcy. Much of debate surrounding marketplace lending platforms centers on whether they help or harm consumer welfare (Danisewicz and Elard, 2018; Cornaggia et al., 2018; Cumming et al., 2022). Our research demonstrates a hitherto neglected unintended effect of the expanding Fintech sector on banks' funding costs. While Fintech appears to have disrupted the deposit market, marketplace lenders remain relatively small which limits the extent of their encroachment into deposit markets. However, it appears reasonable that these effects may strengthen through time as marketplaces originate larger volumes of credit (Thakor (2020)), in which case the Fintech sector may influence monetary policy and macroprudential decision. Given the widespread ramifications of bank funding costs on stability within the sector, bank regulators will have to incorporate Fintech developments into their assessments of financial institutions' health. Exploring these issues is an exciting avenue for future research.

# Tables

Table 1: Variable Descriptions

| Variable | Description |
|---|---|
| Fintech index | An ordinal variable that takes the value of 2 if individuals in state $s$ at quarter $t$ are allowed to invest in both Lending Club and prosper, 1 if they can invest in either of the two platforms, and zero if investors are prohibited from investing in either platform |
| Cost of deposits | The ratio of total deposit interest expenses to total deposits (in natural logarithms) for bank $b$ in state $s$ in quarter $t$ |
| Av. APY | Average annualised percentage yield of deposits for bank $b$ in state $s$ in quarter $t$ |
| Av. Rate | Average quoted rate of deposits for bank $b$ in state $s$ in quarter $t$ |
| APY-CD | Average annualised percentage yield of CD for bank $b$ in state $s$ in quarter $t$ |
| APY-IC | Average annualised percentage yield of Checking accounts for bank $b$ in state $s$ in quarter $t$ |
| APY-IRA | Average annualised percentage yield of fixed and variable interest rates account for bank $b$ in state $s$ in quarter $t$ |
| APY-SAV | Average annualised percentage yield of saving account for bank $b$ in state $s$ in quarter $t$ |
| Rate-CD | Average quoted rate of CD for bank $b$ in state $s$ in quarter $t$ |
| Rate-IC | Average quoted rate of Checking accounts for bank $b$ in state $s$ in quarter $t$ |
| Rate-IRA | Average quoted rate of fixed and variable interest rates account for bank $b$ in state $s$ in quarter $t$ |
| Rate-SAV | Average quoted rate of saving account for bank $b$ in state $s$ in quarter $t$ |
| Deposit share | The ratio of deposits to total liabilities for bank $b$ in state $s$ in quarter $t$ |
| Insured deposits | Insured deposits (in natural logarithms) for bank $b$ in state $s$ in quarter $t$ |
| Uninsured deposits | Uninsured deposits (in natural logarithms) for bank $b$ in state $s$ in quarter $t$ |
| Income growth rate | The annual rate of per capita income growth in state $s$ |
| Population | The annual natural logarithm of population in state $s$ |
| Establishment per capita | The annual number of establishments per capita (in natural logarithms) in state $s$ |
| Unemployment rate | The unemployment rate in state $s$ in quarter $t$ |
| Bank size | The natural logarithm of total assets for bank $b$ in state $s$ in quarter $t$ |
| Capital ratio | The ratio of total assets minus total liabilities to total assets for bank $b$ in state $s$ in quarter $t$ (in natural logarithms) |
| Branches | The number of branches belonging to bank $b$ in state $s$ in quarter $t$ in natural logarithms) |
| Multistate | A dummy variable equal to 1 if bank $b$ operates branches in more than one state in year $t$, 0 otherwise |
| Leverage | The ratio of total liabilities to total assets for bank $b$ in state $s$ in quarter $t$ |
| ROA | The ratio of net profit to total assets for bank $b$ in state $s$ in quarter $t$ |
| ROE | The ratio of net profit to total shareholders' equity for bank $b$ in state $s$ in quarter $t$ |
| $\sigma_{ROA}$ | 12 quarter rolling standard deviation of RoA for bank $b$ in state $s$ in quarter $t$ |
| $\sigma_{ROE}$ | 12 quarter rolling standard deviation of RoE for bank $b$ in state $s$ in quarter $t$ |
| Z-score | (ROA+Capital Ratio)/$\sigma_{ROA}$ for bank $b$ in state $s$ in quarter $t$ |
| Non-deposit cost | The ratio of non-deposit interest expenses to non-deposit liabilities for bank $b$ in state $s$ in quarter $t$ |
| Market share | The ratio of deposits in bank $b$ in state $s$ in quarter $t$ to total deposits held by banks in state $s$ in quarter $t$ |
| HHI index | The Herfindahl–Hirschman index of banks' deposit market share in state $s$ in quarter $t$ |
| Branch closure | A dummy equal to 1 if a bank closes a branch in year $t$, 0 otherwise |
| Branch opening | A dummy equal to 1 if a bank opens a branch in year $t$, 0 otherwise |
| Exit | A dummy equal to 1 if a bank exits in year $t$, 0 otherwise |
| Entry | A dummy equal to 1 if a bank enters in year $t$, 0 otherwise |
| FolioFn | A dummy variable equal to 1 if investing through FolioFn is allowed in state $s$ during quarter $t$ |
| VC amount | VC investment funding per capita in state $s$ during quarter $t$ |
| VC deals | The number of VC deals per capita in state $s$ during quarter $t$ |
| Corporate tax rate | The top marginal corporate tax rate in state $s$ during quarter $t$ |
| Housing price index | The FHFA house price index in state $s$ during quarter $t$ |
| Auto delinquency rate | The share of auto loans that are 90+ days in arrears in state $s$ during quarter $t$ |
| Credit card delinquency rate | The share of credit card loans that are 90+ days in arrears in state $s$ during quarter $t$ |
| Mortgage delinquency rate | The share of mortgage loans that are 90+ days in arrears in state $s$ during quarter $t$ |
| Student loan delinquency rate | The share of student loans that are 90+ days in arrears in state $s$ during quarter $t$ |

Table 2: Descriptive Statistics

| | Mean | SD | Median | P25 | P75 |
|---|---|---|---|---|---|
| Fintech index | 0.8632 | 0.8930 | 0.0000 | 1.0000 | 2.0000 |
| Av. funding rate | 0.0108 | 0.0080 | 0.0042 | 0.0081 | 0.0160 |
| Av. rate (CD) | 0.0196 | 0.0138 | 0.0082 | 0.0149 | 0.0292 |
| Av. rate (IC) | 0.0031 | 0.0037 | 0.0010 | 0.0020 | 0.0043 |
| Av. rate (MM) | 0.0083 | 0.0078 | 0.0025 | 0.0053 | 0.0120 |
| Av. rate (SAV) | 0.0043 | 0.0041 | 0.0010 | 0.0025 | 0.0060 |
| Av. APY | 0.0109 | 0.0081 | 0.0042 | 0.0082 | 0.0161 |
| Av. APY(CD) | 0.0196 | 0.0138 | 0.0082 | 0.0149 | 0.0292 |
| Av. APY(IC) | 0.0031 | 0.0037 | 0.0010 | 0.0020 | 0.0043 |
| Av. APY(MM) | 0.0083 | 0.0078 | 0.0025 | 0.0053 | 0.0120 |
| Av. APY(SAV) | 0.0043 | 0.0041 | 0.0010 | 0.0025 | 0.0060 |
| Platform rate | 0.1527 | 0.0230 | 0.1315 | 0.1528 | 0.1733 |
| Av. loan rate | 0.0690 | 0.0208 | 0.0542 | 0.0675 | 0.0812 |
| Av. spread (Av. loan rate- Av. funding rate) | 0.0582 | 0.0189 | 0.0453 | 0.0559 | 0.0683 |
| Av. platform spread (Av. platform rate- Av. funding rate) | 0.1423 | 0.0231 | 0.1238 | 0.1392 | 0.1651 |
| Deposit growth | 0.0338 | 0.1636 | -0.0420 | 0.0187 | 0.0863 |
| Bank size | 12.4813 | 1.2263 | 11.7810 | 12.3775 | 12.8743 |
| Branches | 2.2616 | 1.0545 | 1.6094 | 2.0794 | 2.7726 |
| Leverage | 0.8948 | 0.0399 | 0.8863 | 0.9029 | 0.9153 |
| ROA | 0.0053 | 0.0077 | 0.0023 | 0.0048 | 0.0086 |
| ROE | 0.0060 | 0.0094 | 0.0025 | 0.0054 | 0.0097 |
| $\sigma_{ROA}$ | 0.0056 | 0.0040 | 0.0031 | 0.0043 | 0.0065 |
| $\sigma_{ROE}$ | 0.0068 | 0.0066 | 0.0034 | 0.0048 | 0.0074 |
| Non-deposit cost | 0.0213 | 0.0134 | 0.0110 | 0.0192 | 0.0282 |
| Market share | 0.0067 | 0.0138 | 0.0012 | 0.0029 | 0.0062 |
| Z-score | 26.0561 | 13.1827 | 16.2903 | 24.9007 | 34.6408 |
| Market share | 0.0055 | 0.0163 | 0.0014 | 0.0005 | 0.0036 |
| HHI | 0.1569 | 0.1638 | 0.0476 | 0.0855 | 0.2343 |
| Income growth | 0.0410 | 0.0296 | 0.0290 | 0.0429 | 0.0590 |
| Unemployment | 0.0603 | 0.0209 | 0.0456 | 0.0537 | 0.0739 |
| Population growth | 0.0078 | 0.0065 | 0.0028 | 0.0064 | 0.0117 |
| Establishment per capita | 0.0250 | 0.0035 | 0.0222 | 0.0244 | 0.0269 |
| Branch opening | 0.0643 | 0.2454 | 0.0000 | 0.0000 | 0.0000 |
| Branch closure | 0.0422 | 0.2011 | 0.0000 | 0.0000 | 0.0000 |
| Exit | 0.0064 | 0.0799 | 0.0000 | 0.0000 | 0.0000 |
| Entry | 0.0003 | 0.0165 | 0.0000 | 0.0000 | 0.0000 |
| Corporate tax rate | 0.0350 | 0.0381 | 0.0000 | 0.0000 | 0.0697 |
| Usury rate | 0.2168 | 0.1732 | 0.0800 | 0.1600 | 0.4500 |
| Auto delinquency | 0.0341 | 0.0137 | 0.0234 | 0.0307 | 0.0419 |
| Credit card delinquency | 0.0890 | 0.0256 | 0.0706 | 0.0844 | 0.1017 |
| Mortgage delinquency | 0.0305 | 0.0286 | 0.0132 | 0.0218 | 0.0383 |
| Student loan delinquency | 0.0926 | 0.0280 | 0.0716 | 0.0897 | 0.1149 |
| Bank-branch quarter observations | 208,171 | - | - | - | - |

## Table 3: Timing of Restriction Removal across States

| State | Lending Club | Prosper |
|---|---|---|
| Alabama | 2015Q4 | - |
| Alaska | - | 2010Q3 |
| Arizona | 2015Q2 | - |
| Arkansas | 2015Q3 | - |
| California | 2008Q4 | 2007 Q1 |
| Colorado | 2008Q4 | 2007 Q1 |
| Connecticut | 2008Q4 | 2007 Q1 |
| DC | 2015Q4 | 2007 Q1 |
| Delaware | 2008Q4 | 2007 Q1 |
| Florida | 2008Q4 | 2007 Q1 |
| Georgia | 2008Q4 | 2007 Q1 |
| Hawaii | 2008Q4 | 2007 Q1 |
| Idaho | 2008Q4 | 2007 Q1 |
| Illinois | 2008Q4 | 2007 Q1 |
| Indiana | 2015Q3 | 2015Q3 |
| Iowa | 2015Q3 | - |
| Kansas | 2015Q3 | - |
| Kentucky | 2015Q4 | - |
| Louisiana | 2008 Q4 | 2007 Q1 |
| Maine | 2009Q3 | 2007 Q1 |
| Maryland | 2016Q1 | - |
| Massachusetts | 2014Q4 | - |
| Michigan | 2015Q4 | 2014 Q1 |
| Minnesota | 2008Q4 | 2007 Q1 |
| Mississippi | 2008Q4 | 2007 Q1 |
| Montana | 2008Q4 | 2007 Q1 |
| Nebraska | 2015Q3 | - |
| Nevada | 2008Q4 | 2007 Q1 |
| New Hampshire | 2008Q4 | 2007 Q1 |
| New Jersey | 2016Q1 | - |
| New Mexico | - | - |
| New York | 2008Q4 | 2007 Q1 |
| North Carolina | 2010Q4 | - |
| North Dakota | 2016Q1 | - |
| Ohio | - | - |
| Oklahoma | 2015Q3 | - |
| Oregon | 2016Q1 | 2007 Q1 |
| Pennsylvania | - | - |
| Rhode Island | 2008Q4 | 2007 Q1 |
| South Carolina | 2008Q4 | 2007 Q1 |
| South Dakota | 2008Q4 | 2007 Q1 |
| Tennessee | - | 2019Q1 |
| Texas | 2015Q2 | 2019 Q1 |
| Utah | 2008Q4 | 2012Q4 |
| Vermont | 2014Q3 | 2012 Q1 |
| Virginia | 2008Q4 | 2007 Q1 |
| Washington | 2016Q1 | 2007 Q1 |
| West Virginia | 2008Q4 | 2012 Q1 |
| Wisconsin | 2008Q4 | 2007 Q1 |
| Wyoming | 2008Q4 | 2007 Q1 |

Notes: This table reports the quarter when a state security regulator removed restrictions on investing through Lending Club and Prosper by individuals and businesses in the state. – indicates that a state security regulator has not removed investing restrictions on a marketplace.

## Table 4: Identifying Assumptions Tests

|  | 1 | 2 | 3 | 4 | 5 | 6 | 7 | 8 |
|---|---|---|---|---|---|---|---|---|
|  | Funding Rate (FR) | | | | Annualized Percentage Yield (APY) | | | |
| Fintech index$_{st-1}$ | 0.0094 (0.0364) | | | | 0.0095 (0.0372) | | | |
| Fintech index$_{st-2}$ | | 0.0066 (0.0387) | | | | 0.0063 (0.0393) | | |
| Fintech index$_{st-3}$ | | | 0.0132 (0.0388) | | | | 0.0127 (0.0393) | |
| Fintech index$_{st-4}$ | | | | 0.0166 (0.0306) | | | | 0.0139 (0.0350) |
| Income growth | -0.0071* (0.0037) | -0.0076** (0.0037) | -0.0071* (0.0037) | -0.0071* (0.0037) | -0.0076** (0.0037) | -0.0071* (0.0037) | -0.0071* (0.0037) | -0.0071* (0.0037) |
| Population | -0.0131 (0.0132) | -0.0132 (0.0135) | -0.0127 (0.0132) | -0.0131 (0.0132) | -0.0132 (0.0135) | -0.0127 (0.0132) | -0.0127 (0.0132) | -0.0131 (0.0132) |
| Establishments per capita | -0.3217** (0.1601) | -0.3356** (0.1614) | -0.3180* (0.1602) | -0.3217** (0.1601) | -0.3356** (0.1614) | -0.3180* (0.1602) | -0.3180* (0.1602) | -0.3217** (0.1601) |
| Unemployment Rate | -0.0336** (0.0142) | -0.0343** (0.0140) | -0.0335** (0.0142) | -0.0336** (0.0140) | -0.0343** (0.0143) | -0.0335** (0.0142) | -0.0335** (0.0143) | -0.0336** (0.0142) |
| Bank FE | Yes | Yes | Yes | Yes | Yes | Yes | Yes | Yes |
| Quarter-Year FE | Yes | Yes | Yes | Yes | Yes | Yes | Yes | Yes |
| Adj. $R^2$ | 0.8590 | 0.8592 | 0.8595 | 0.8590 | 0.8592 | 0.8595 | 0.8592 | 0.8595 |
| Observations | 98,418 | 98,418 | 98,418 | 98,418 | 98,418 | 98,418 | 98,418 | 98,418 |

Notes: This table reports estimates of equation (2). The dependent variable in columns is average funding rate (and Annual percentage yield). Fintech index$_{st-1}$ is a dummy variable equal to 1 in the quarter prior to the removal of investment restrictions on either Lending Club or Prosper in state $s$, 0 otherwise. Fintech index$_{st-2}$ is a dummy variable equal to 1 two quarters prior to the removal of investment restrictions on either Lending Club or Prosper in state $s$, 0 otherwise. Fintech index$_{st-3}$ is a dummy variable equal to 1 three quarters prior to the removal of investment restrictions on either Lending Club or Prosper in state $s$, 0 otherwise. Fintech index$_{st-4}$ is a dummy variable equal to 1 four quarters prior to the removal of investment restrictions on either Lending Club or Prosper in state $s$, 0 otherwise. The standard errors are two-way clustered at the bank and quarter-year levels and are reported in parentheses.

## Table 5: Pre-treatment Characteristics

| | Treatment | Control | Diff | t-stat |
|---|---|---|---|---|
| Size | 12.5216 | 12.5352 | -0.0135 | -0.8537 |
| Branches | 2.2278 | 2.2074 | 0.0203 | 1.4828 |
| Leverage | 0.8941 | 0.8949 | -0.0008 | -1.5084 |
| ROA | 0.0045 | 0.0045 | 0.0000 | -0.5122 |
| ROE | 0.0050 | 0.0051 | 0.0000 | -0.2301 |
| $\sigma_{ROA}$ | 0.0058 | 0.0059 | 0.0000 | -0.8929 |
| $\sigma_{ROE}$ | 0.0071 | 0.0070 | 0.0000 | 0.2207 |
| Z-score | 31.1070 | 31.4540 | -0.3470 | -0.66 |
| Non-deposit cost | 0.0209 | 0.0208 | 0.0001 | 0.5832 |

Notes: This table reports estimates from *t*-tests that test equality in the mean pre-treatment values of bank characteristics between control and treated banks. Variable definitions are reported in Table 1. Control denotes banks are those headquartered in states at *t*-1 that impose investment restrictions on Lending Club and Prosper at time *t*-1 and *t*. Treatment denotes banks are those headquartered in states at *t*-1 that impose investment restrictions on Lending Club and Prosper at time *t*-1 but not at time *t*. Control (Treatment) is the mean value of the variable among control (treated) banks. Difference is equal to Control – Treatment. *t*-statistic is the *t*-statistic from a *t*-test of equality between Control and Treatment.

## Table 6: Lending Test

| Dependent variable: | 1<br>Marketplace lending (ln) |
|---|---|
| Fintech Index | 0.2831** |
|  | (0.1313) |
| Income growth | -0.0408 |
|  | (0.0337) |
| Population | 2.7888 |
|  | (2.5679) |
| Establishment per capita | -3.0444 |
|  | (4.2149) |
| Unemployment Rate | 0.3612*** |
|  | (0.0718) |
| State FE | Yes |
| Quarter x Year FE | Yes |
| Adj. $R^2$ | 0.8414 |
| Observations | 2,981 |

Notes: This table reports estimates of $L_{st} = \alpha + \beta Fintech\ Index_{st} + \gamma X_{st} + \varphi_s + \varphi_t + \varepsilon_{st}$ where $L_{st}$ is the natural logarithm of total lending by Lending Club and Prosper in state $s$ during quarter-year $t$; $X_{st}$ is a vector of time-varying state-level control variables; $\varphi_s$ and $\varphi_t$ denote state and quarter-year fixed effects, respectively; $\varepsilon_{ist}$ is the error term. Variable definitions are reported in Table 1. The standard errors are two-way clustered at the state and quarter-year levels and are reported in parentheses. *, **, and *** indicates statistical significance at the 10%, 5%, and 1% levels, respectively.

## Table 7: Fintech Competition and Banks' Cost of Deposits

|  | 1 | 2 | 3 | 4 |
|---|---|---|---|---|
| Panel A: Funding rate |  |  |  |  |
| Fintech index | 0.0297*** | 0.0299*** | 0.0299*** | 0.0158** |
|  | (0.0034) | (0.0034) | (0.0034) | (0.0072) |
| Income growth | -0.0060*** | -0.0060*** | -0.0060*** | -0.0062*** |
|  | (0.0008) | (0.0008) | (0.0008) | (0.0009) |
| Population | 0.0606*** | 0.0606*** | 0.0606*** | 0.0610** |
|  | (0.0093) | (0.0094) | (0.0094) | (0.0248) |
| Establishments per capita | 0.7892*** | 0.7891*** | 0.7893*** | 0.7753*** |
|  | (0.0679) | (0.0678) | (0.0678) | (0.1452) |
| Unemployment rate | -0.0089*** | -0.0089*** | -0.0089*** | -0.0088*** |
|  | (0.0021) | (0.0021) | (0.0021) | (0.0031) |
| Fintech index × bank size |  | -0.0007* |  |  |
|  |  | (0.0004) |  |  |
| Fintech index × branches |  |  | -0.0030** |  |
|  |  |  | (0.0014) |  |
| Lending (ln) | | | | -0.0018* |
| | | | | (0.0009) |
| Fintech index × lending (ln) | | | | 0.0010** |
| | | | | (0.0005) |
| Bank × Quarter × Year FE | Yes | Yes | Yes | Yes |
| Branch FE | Yes | Yes | Yes | Yes |
| Adj. $R^2$ | 0.919 | 0.919 | 0.919 | 0.919 |
| Observations | 208,171 | 208,171 | 208,171 | 208,171 |
| Panel B: APY | | | | |
| Fintech index | 0.0297*** | 0.0299*** | 0.0299*** | 0.0157** |
| | (0.0034) | (0.0035) | (0.0034) | (0.0073) |
| Income growth | -0.0060*** | -0.0060*** | -0.0060*** | -0.0062*** |
| | (0.0008) | (0.0008) | (0.0008) | (0.0009) |
| Population | 0.0615*** | 0.0615*** | 0.0615*** | 0.0620** |
| | (0.0094) | (0.0094) | (0.0094) | (0.0249) |
| Establishments per capita | 0.7979*** | 0.7977*** | 0.7980*** | 0.7839*** |
| | (0.0681) | (0.0680) | (0.0680) | (0.1455) |
| Unemployment rate | -0.0090*** | -0.0090*** | -0.0090*** | -0.0089*** |
| | (0.0021) | (0.0021) | (0.0021) | (0.0032) |
| Fintech index × bank size | | -0.0007* | | |
| | | (0.0004) | | |
| Fintech index × branches | | | -0.0030** | |
| | | | (0.0014) | |
| Lending (ln) | | | | -0.0019** |
| | | | | (0.0009) |
| Fintech index × lending (ln) | | | | 0.0010** |
| | | | | (0.0005) |
| Bank × Quarter × Year FE | Yes | Yes | Yes | Yes |
| Branch FE | Yes | Yes | Yes | Yes |
| Adj. $R^2$ | 0.920 | 0.920 | 0.920 | 0.919 |
| Observations | 208,171 | 208,171 | 208,171 | 208,171 |

Notes: This table reports estimates of equations (1) and (2). The dependent variable is the cost of deposits. Variable definitions are reported in Table 1. The standard errors are two-way clustered at the bank and quarter-year levels and are reported in parentheses. *, **, and *** indicates statistical significance at the 10%, 5%, and 1% levels, respectively.

### Table 8: Deposit Product-level Effects and Spread

| | 1 | 2 | 3 | 4 | 5 | 6 | 7 | 8 | 9 | 10 | 11 |
|---|---|---|---|---|---|---|---|---|---|---|---|
| Dependent variable | Funding rate | | | | Annual percentage yield | | | | Spread (Platform rate-av. Deposit rate) | Spread (Av. loan rate- Av. Deposit rate) | Funding rate |
| Account type | IC | MM | SAV | CD | IC | MM | SAV | CD | - | - | - |
| Fintech index | 0.0293*** | 0.0272*** | 0.0255*** | 0.0251*** | 0.0293*** | 0.0272*** | 0.0255*** | 0.0251*** | 0.0370*** | 0.0178*** | 0.0296*** |
| | (0.0070) | (0.0055) | (0.0051) | (0.0030) | (0.0070) | (0.0055) | (0.0051) | (0.0030) | (0.0021) | (0.0038) | (0.0034) |
| Income growth | -0.0035** | -0.0116*** | -0.0051*** | -0.0025*** | -0.0035** | -0.0116*** | -0.0051*** | -0.0025*** | 0.0067*** | 0.0066*** | -0.0060*** |
| | (0.0015) | (0.0014) | (0.0007) | | (0.0015) | (0.0014) | (0.0007) | (0.0009) | (0.0025) | (0.0008) | |
| Population | 0.1593*** | 0.0219 | 0.0294 | 0.0396*** | 0.1593*** | 0.0219 | 0.0294 | 0.0396*** | 0.0063*** | 0.0237*** | 0.0600*** |
| | (0.0298) | (0.0202) | (0.0220) | (0.0085) | (0.0298) | (0.0202) | (0.0220) | (0.0085) | (0.0020) | (0.0038) | (0.0094) |
| Establishments per capita | 1.0143*** | 0.4993*** | 0.9506*** | 0.9612*** | 1.0143*** | 0.4993*** | 0.9506*** | 0.9612*** | 0.0789*** | 0.1642*** | 0.7864*** |
| | (0.1472) | (0.1418) | (0.1151) | (0.0596) | (0.1472) | (0.1418) | (0.1151) | (0.0596) | (0.0163) | (0.0304) | (0.0678) |
| Unemployment rate | 0.0012 | 0.0032 | -0.0068** | -0.0080*** | 0.0012 | 0.0032 | -0.0068** | -0.0080*** | 0.0107*** | -0.0032 | -0.0090*** |
| | (0.0038) | (0.0033) | (0.0031) | (0.0018) | (0.0038) | (0.0033) | (0.0031) | (0.0018) | (0.0017) | (0.0036) | (0.0021) |
| Ln (Loan Rate) | | | | | | | | | | | 0.0658** |
| | | | | | | | | | | | (0.0291) |
| Bank × Quarter × Year FE | Yes | Yes | Yes | Yes | Yes | Yes | Yes | Yes | Yes | Yes | Yes |
| Branch FE | Yes | Yes | Yes | Yes | Yes | Yes | Yes | Yes | Yes | Yes | Yes |
| Adj. $R^2$ | 0.714 | 0.803 | 0.866 | 0.916 | 0.714 | 0.803 | 0.866 | 0.916 | 0.873 | 0.297 | 0.919 |
| Observations | 208,171 | 208,171 | 208,171 | 208,171 | 208,171 | 208,171 | 208,171 | 208,171 | 208,171 | 208,171 | 208,171 |

Notes: This table reports estimates of equations (2). The dependent variable is the cost of deposits. Variable definitions are reported in Table 1. The standard errors are two-way clustered at the bank and quarter-year levels and are reported in parentheses. *, **, and *** indicates statistical significance at the 10%, 5%, and 1% levels, respectively.

## Table 9: Bank Conditions and Debtholder Monitoring

| | 1 | 2 | 3 | 4 | 5 | 6 | 7 |
|---|---|---|---|---|---|---|---|
| **Panel A: Funding rate** | | | | | | | |
| Fintech index | 0.0292*** | 0.0298*** | 0.0298*** | 0.0293*** | 0.0293*** | 0.0291*** | 0.0293*** |
| | (0.0038) | (0.0037) | (0.0037) | (0.0037) | (0.0037) | (0.0038) | (0.0038) |
| Fintech index × Z-score | 0.0004 | | | | | | |
| | (0.0006) | | | | | | |
| Fintech index × ROA | | -0.0001 | | | | | |
| | | (0.0006) | | | | | |
| Fintech index × ROE | | | -0.0001 | | | | |
| | | | (0.0006) | | | | |
| Fintech index × $\sigma_{ROA}$ | | | | 0.0003 | | | |
| | | | | (0.0006) | | | |
| Fintech index × $\sigma_{ROE}$ | | | | | 0.0003 | | |
| | | | | | (0.0006) | | |
| Fintech index × Leverage | | | | | | 0.0005 | |
| | | | | | | (0.0006) | |
| Fintech index × Non-deposit cost | | | | | | | 0.0004 |
| | | | | | | | (0.0006) |
| Control variables | Yes | Yes | Yes | Yes | Yes | Yes | Yes |
| Bank × Quarter × Year FE | Yes | Yes | Yes | Yes | Yes | Yes | Yes |
| Branch FE | Yes | Yes | Yes | Yes | Yes | Yes | Yes |
| Adj. $R^2$ | 0.919 | 0.919 | 0.919 | 0.919 | 0.919 | 0.919 | 0.919 |
| Observations | 208,171 | 208,171 | 208,171 | 208,171 | 208,171 | 208,171 | 208,171 |
| **Panel B: APY** | | | | | | | |
| Fintech index | 0.0292*** | 0.0298*** | 0.0298*** | 0.0294*** | 0.0294*** | 0.0291*** | 0.0293*** |
| | (0.0038) | (0.0037) | (0.0037) | (0.0038) | (0.0038) | (0.0038) | (0.0038) |
| Fintech index × Z-score | 0.0004 | | | | | | |
| | (0.0006) | | | | | | |
| Fintech index × ROA | | -0.0001 | | | | | |
| | | (0.0006) | | | | | |
| Fintech index × ROE | | | -0.0001 | | | | |
| | | | (0.0006) | | | | |
| Fintech index × $\sigma_{ROA}$ | | | | 0.0003 | | | |
| | | | | (0.0006) | | | |
| Fintech index × $\sigma_{ROE}$ | | | | | 0.0003 | | |
| | | | | | (0.0006) | | |
| Fintech index × Leverage | | | | | | 0.0005 | |
| | | | | | | (0.0006) | |
| Fintech index × Non-deposit cost | | | | | | | 0.0003 |
| | | | | | | | (0.0006) |
| Control variables | Yes | Yes | Yes | Yes | Yes | Yes | Yes |
| Bank × Quarter × Year FE | Yes | Yes | Yes | Yes | Yes | Yes | Yes |
| Branch FE | Yes | Yes | Yes | Yes | Yes | Yes | Yes |
| Adj. $R^2$ | 0.920 | 0.920 | 0.920 | 0.920 | 0.920 | 0.920 | 0.920 |
| Observations | 208,171 | 208,171 | 208,171 | 208,171 | 208,171 | 208,171 | 208,171 |

Notes: This table reports estimates of equation (2). Variable definitions are reported in Table 1. The unreported control variable are income growth, population, establishments per capita, unemployment rate, bank size, capital ratio, and branches. The standard errors are two-way clustered at the bank and quarter-year levels and are reported in parentheses. *, **, and *** indicates statistical significance at the 10%, 5%, and 1% levels, respectively.

## Table 10: Market Power and Competition

| | 1 | 2 | 3 | 4 | 5 | 6 |
|---|---|---|---|---|---|---|
| **Panel A: Funding rate** | | | | | | |
| Fintech index | 0.0292*** | 0.0290*** | 0.0298*** | 0.0299*** | 0.0297*** | 0.0297*** |
| | (0.0037) | (0.0037) | (0.0034) | (0.0034) | (0.0034) | (0.0034) |
| Fintech index × Market Share | 0.0004 | | | | | |
| | (0.0006) | | | | | |
| Fintech index × HHI | | 0.0001 | | | | |
| | | (0.0003) | | | | |
| Fintech index × Branch closure | | | -0.0019 | | | |
| | | | (0.0071) | | | |
| Fintech index × Branch opening | | | | -0.0054 | | |
| | | | | (0.0050) | | |
| Fintech index × Bank exit | | | | | -0.0181 | |
| | | | | | (0.0118) | |
| Fintech index × Bank entry | | | | | | 0.0000 |
| | | | | | | (0.0001) |
| Control variables | Yes | Yes | Yes | Yes | Yes | Yes |
| Bank × Quarter × Year FE | Yes | Yes | Yes | Yes | Yes | Yes |
| Branch FE | Yes | Yes | Yes | Yes | Yes | Yes |
| Adj. $R^2$ | 0.919 | 0.919 | 0.919 | 0.919 | 0.919 | 0.919 |
| Observations | 208,171 | 208,171 | 208,171 | 208,171 | 208,171 | 208,171 |
| **Panel B: APY** | | | | | | |
| Fintech index | 0.0292*** | 0.0290*** | 0.0298*** | 0.0299*** | 0.0297*** | 0.0297*** |
| | (0.0038) | (0.0037) | (0.0034) | (0.0034) | (0.0034) | (0.0034) |
| Fintech index × Market Share | 0.0004 | | | | | |
| | (0.0006) | | | | | |
| Fintech index × HHI | | 0.0001 | | | | |
| | | (0.0003) | | | | |
| Fintech index × Branch closure | | | -0.0019 | | | |
| | | | (0.0071) | | | |
| Fintech index × Branch opening | | | | -0.0057 | | |
| | | | | (0.0051) | | |
| Fintech index × Bank exit | | | | | -0.0184 | |
| | | | | | (0.0118) | |
| Fintech index × Bank entry | | | | | | 0.0000 |
| | | | | | | (0.0001) |
| Control variables | Yes | Yes | Yes | Yes | Yes | Yes |
| Bank × Quarter × Year FE | Yes | Yes | Yes | Yes | Yes | Yes |
| Branch FE | Yes | Yes | Yes | Yes | Yes | Yes |
| Adj. $R^2$ | 0.920 | 0.920 | 0.920 | 0.920 | 0.920 | 0.920 |
| Observations | 208,171 | 208,171 | 208,171 | 208,171 | 208,171 | 208,171 |

Notes: This table reports estimates of equation (2). Variable definitions are reported in Table 1. The unreported control variable are income growth, population, establishments per capita, unemployment rate, bank size, capital ratio, and branches. The standard errors are two-way clustered at the bank and quarter-year levels and are reported in parentheses. *, **, and *** indicates statistical significance at the 10%, 5%, and 1% levels, respectively.

## Table 11: Industry Dynamics and Survivorship Bias

| | 1 | 2 | 3 | 4 |
|---|---|---|---|---|
| Dependent variable | Funding rate | | APY | |
| Sample | Ex. failed banks | Ex. entrants | Ex. failed banks | Ex. entrants |

| | (1) | (2) | (3) | (4) |
|---|---|---|---|---|
| Fintech index | 0.0298*** | 0.0297*** | 0.0298*** | 0.0297*** |
| | (0.0034) | (0.0034) | (0.0034) | (0.0034) |
| Income growth | -0.0060*** | -0.0060*** | -0.0060*** | -0.0060*** |
| | (0.0008) | (0.0008) | (0.0008) | (0.0008) |
| Population | 0.0597*** | 0.0605*** | 0.0606*** | 0.0615*** |
| | (0.0093) | (0.0094) | (0.0094) | (0.0094) |
| Establishment per capita | 0.7803*** | 0.7892*** | 0.7889*** | 0.7978*** |
| | (0.0680) | (0.0679) | (0.0682) | (0.0681) |
| Unemployment Rate | -0.0088*** | -0.0089*** | -0.0089*** | -0.0090*** |
| | (0.0021) | (0.0021) | (0.0021) | (0.0021) |
| Bank × Quarter × Year FE | Yes | Yes | Yes | Yes |
| Branch FE | Yes | Yes | Yes | Yes |
| Adj. $R^2$ | 0.920 | 0.919 | 0.920 | 0.920 |
| Observations | 207,279 | 208,133 | 207,278 | 208,132 |

Notes: This table reports estimates of equation (2). Variable definitions are reported in Table 1. The standard errors are two-way clustered at the bank and quarter-year levels and are reported in parentheses. *, **, and *** indicates statistical significance at the 10%, 5%, and 1% levels, respectively.

# Figures

Figure 1: Dynamic DiD (CSDID) based on Callaway and Sant'Anna (2021)

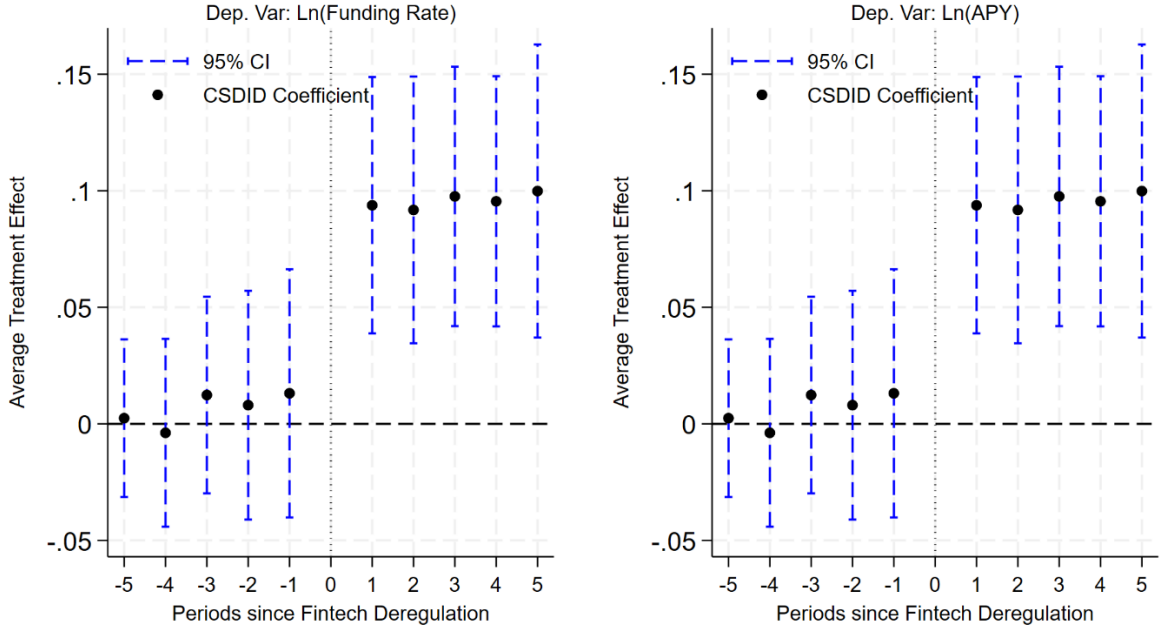

Notes: This figure illustrates the coefficient estimates from an interaction weighted CSDID based on Callaway and Sant'Anna (2021). We normalize the quarter of deregulation to 0 and report estimates for the 5 quarters on either side. The black dots denote the average treatment on the treated effect, and the dotted blue lines illustrate the corresponding lower and upper bounds of a 95% confidence interval.

Figure 2: Placebo Simulation Distribution

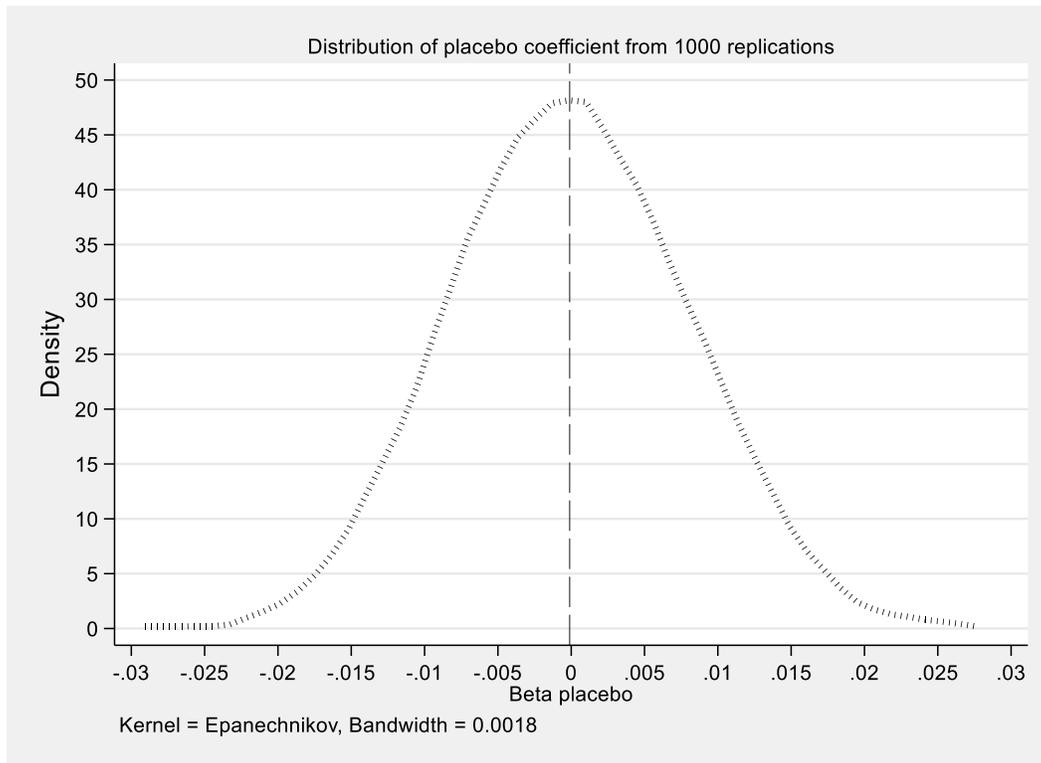

Notes: The figure shows the distribution of placebo coefficient from 1000 replications. 44 out of 1000 simulations (<5%) reject null hypothesis of zero placebo coefficient.

# Online Appendix

-   FOR ONLINE PUBLICATION ONLY   -

## Table 1.A: Deregulation Tests

| | 1 | 2 |
|---|---|---|
| Estimator | Cox PH | Weibull |
| Dependent variable: deregulate dummy | | |
| Population | -0.0142 | -0.0167 |
| | (0.0524) | (0.0734) |
| Unemployment rate | 0.0748 | 0.1060 |
| | (0.0512) | (0.0791) |
| Auto delinquency rate | 0.0058 | -0.0219 |
| | (0.0596) | (0.0899) |
| Credit card delinquency rate | -0.0765 | -0.0692 |
| | (0.0489) | (0.0731) |
| Mortgage delinquency rate | 0.0024 | -0.0085 |
| | (0.0299) | (0.0428) |
| Student delinquency rate | 0.0093 | 0.0073 |
| | (0.0243) | (0.0351) |
| Corporate tax rate | 0.0014 | -0.0008 |
| | (0.0191) | (0.0276) |
| Deposit rate | 0.1578 | 1.1196 |
| | (0.3762) | (0.5804) |
| Z-score | -0.0003 | -0.0006 |
| | (0.0060) | (0.0090) |
| Bank size | 0.0077 | 0.0092 |
| | (0.0152) | (0.0225) |
| Quarter x Year FE | Yes | Yes |
| Observations | 1,690 | 1,690 |

Notes: This table reports estimates of equation (2). We remove all observations from state *s* following the quarter after the removal of investing restrictions. Variable definitions are reported in Table 1. The standard errors two-way clustered at the state and quarter-year levels and are reported in parentheses. *, **, and *** indicates statistical significance at the 10%, 5%, and 1% levels, respectively.

## Table 2.A: Dynamic Difference-in-Difference Estimation Results

| Period/ Dependent variable | 1<br>Funding Rate | 2<br>APY |
|---|---|---|
| Pre[-5] | 0.0024 | 0.0029 |
|  | (0.0173) | (0.0172) |
| Pre[-4] | -0.0039 | -0.0037 |
|  | (0.0206) | (0.0205) |
| Pre[-3] | 0.0124 | 0.0126 |
|  | (0.0215) | (0.0215) |
| Pre[-2] | 0.0080 | 0.0078 |
|  | (0.0250) | (0.0251) |
| Pre[-1] | 0.0131 | 0.0128 |
|  | (0.0272) | (0.0273) |
| Post[+1] | 0.0938*** | 0.0941*** |
|  | (0.0281) | (0.0281) |
| Post[+2] | 0.0918*** | 0.0915*** |
|  | (0.0292) | (0.0291) |
| Post[+3] | 0.0976*** | 0.0979*** |
|  | (0.0284) | (0.0283) |
| Post[+4] | 0.0955*** | 0.0960*** |
|  | (0.0274) | (0.0274) |
| Post[+5] | 0.0999*** | 0.1010*** |
|  | (0.0321) | (0.0320) |
|  |  |  |
| Pre Avg. | 0.0064 | 0.0065 |
|  | (0.0061) | (0.0061) |
| Post Avg. | 0.0924*** | 0.0928*** |
|  | (0.0214) | (0.0214) |
| Controls | Yes | Yes |
| Bank x Quarter x Year FE | Yes | Yes |
| Branch FE | Yes | Yes |
| Adj. $R^2$ | 0.91 | 0.92 |
| Observations | 206,184 | 206,184 |

Notes: This table reports dynamic DID estimates based on Callaway and Sant'Anna (2021). Variable definitions are reported in Table 1. The unreported control variable are income growth, population, establishments per capita, unemployment rate, bank size, capital ratio, and branches. The standard errors are two-way clustered at the bank and quarter-year levels and are reported in parentheses. *, **, and *** indicates statistical significance at the 10%, 5%, and 1% levels, respectively.

Table 3.A: Heterogeneity Tests

| Dependent variable | 1<br>Funding rate | 2<br>APY |
|---|---|---|
| Panel A: competition | | |
| Fintech index | 0.0447** | 0.0446** |
|  | (0.0216) | (0.0217) |
| Fintech index * (1-HHI) | -0.0166 | -0.0166 |
|  | (0.0166) | (0.0167) |
| Income growth | -0.0060*** | -0.0060*** |
|  | (0.0018) | (0.0018) |
| Population | 0.0608 | 0.0618 |
|  | (0.0375) | (0.0374) |
| Establishment per capita | 0.7939*** | 0.8025*** |
|  | (0.2620) | (0.2623) |
| Unemployment Rate | -0.0090 | -0.0091 |
|  | (0.0075) | (0.0076) |
| (1-HHI) | 0.0174 | 0.0177 |
|  | (0.0193) | (0.0193) |
| Bank x Quarter x Year FE | Yes | Yes |
| Branch FE | Yes | Yes |
| Adj. $R^2$ | 0.9194 | 0.9196 |
| Observations | 208,171 | 208,171 |
| Panel B: urbanization | | |
| Fintech index | 0.0277** | 0.0276** |
|  | (0.0129) | (0.0129) |
| Fintech index * Rural | 0.0067 | 0.0070 |
|  | (0.0100) | (0.0100) |
| Income growth | -0.0059*** | -0.0060*** |
|  | (0.0018) | (0.0018) |
| Population | 0.0611 | 0.0621 |
|  | (0.0372) | (0.0371) |
| Establishment per capita | 0.7950*** | 0.8038*** |
|  | (0.2602) | (0.2605) |
| Unemployment Rate | -0.0088 | -0.0089 |
|  | (0.0075) | (0.0075) |
| Bank x Quarter x Year FE | Yes | Yes |
| Branch FE | Yes | Yes |
| Adj. $R^2$ | 0.9194 | 0.9196 |
| Observations | 208,171 | 208,171 |

Notes: This table reports estimates of equation (1). Variable definitions are reported in Table 1. The standard errors are two-way clustered at the bank and quarter-year levels and are reported in parentheses. *, **, and *** indicates statistical significance at the 10%, 5%, and 1% levels, respectively.

## Table 4.A: Long-run Effects

| Dependent variable | 1<br>ROA | 2<br>ROE | 3<br>Leverage |
|---|---|---|---|
| Fintech Index | -0.0002* | -0.0002 | 0.0009** |
|  | (0.0001) | (0.0002) | (0.0004) |
| Bank size | 0.0012*** | 0.0011 | 0.0292*** |
|  | (0.0005) | (0.0007) | (0.0027) |
| Income growth | 0.0001** | 0.0001** | 0.0001 |
|  | (0.0000) | (0.0000) | (0.0001) |
| Population | 0.0002* | 0.0001 | 0.0005 |
|  | (0.0001) | (0.0001) | (0.0004) |
| Establishment per capita | 0.0002 | -0.0006 | -0.0081*** |
|  | (0.0008) | (0.0010) | (0.0030) |
| Unemployment Rate | -0.0006*** | -0.0008*** | 0.0016*** |
|  | (0.0001) | (0.0001) | (0.0004) |
| Bank FE | Yes | Yes | Yes |
| Quarter x Year FE | Yes | Yes | Yes |
| Adj. $R^2$ | 0.5827 | 0.5635 | 0.7890 |
| Observations | 36073 | 36073 | 36073 |

Notes: This table reports estimates of equation (1) using dependent variables as ROA (column 1), ROE (column 2) and Leverage (column 3). Variable definitions are reported in Table 1. The standard errors are two-way clustered at the bank and quarter-year levels and are reported in parentheses. *, **, and *** indicates statistical significance at the 10%, 5%, and 1% levels, respectively.

Table 5.A: Falsification Tests

| | 1 | 2 | 3 | 4 | 5 |
|---|---|---|---|---|---|
| Panel A: Funding rate | Random Assignment to false treatment | | | | |
| Sample | 50-50 | 30% | 45% | 60% | 75% |
| Placebo | -0.0015 | 0.0009 | -0.0011 | -0.0017 | -0.0013 |
| | (0.0016) | (0.0017) | (0.0016) | (0.0016) | (0.0018) |
| Income growth | -0.0065*** | -0.0065*** | -0.0065*** | -0.0065*** | -0.0065*** |
| | (0.0011) | (0.0011) | (0.0011) | (0.0011) | (0.0011) |
| Population | 0.0895*** | 0.0894*** | 0.0895*** | 0.0895*** | 0.0895*** |
| | (0.0177) | (0.0177) | (0.0177) | (0.0177) | (0.0177) |
| Establishments per capita | 0.8259*** | 0.8257*** | 0.8258*** | 0.8258*** | 0.8257*** |
| | (0.1609) | (0.1610) | (0.1609) | (0.1609) | (0.1610) |
| Unemployment rate | -0.0140*** | -0.0140*** | -0.0140*** | -0.0140*** | -0.0140*** |
| | (0.0033) | (0.0033) | (0.0033) | (0.0033) | (0.0033) |
| Bank × Quarter × Year FE | Yes | Yes | Yes | Yes | Yes |
| Branch FE | Yes | Yes | Yes | Yes | Yes |
| Adj. $R^2$ | 0.944 | 0.944 | 0.944 | 0.944 | 0.944 |
| Observations | 98,951 | 98,951 | 98,951 | 98,951 | 98,951 |
| Panel B: APY | Random Assignment to false treatment | | | | |
| Sample | 50-50 | 30% | 45% | 60% | 75% |
| Placebo | -0.0015 | 0.0009 | -0.0011 | -0.0016 | -0.0012 |
| | (0.0016) | (0.0018) | (0.0016) | (0.0016) | (0.0018) |
| Income growth | -0.0065*** | -0.0065*** | -0.0065*** | -0.0065*** | -0.0065*** |
| | (0.0011) | (0.0011) | (0.0011) | (0.0011) | (0.0011) |
| Population | 0.0920*** | 0.0918*** | 0.0919*** | 0.0920*** | 0.0919*** |
| | (0.0176) | (0.0177) | (0.0176) | (0.0176) | (0.0176) |
| Establishments per capita | 0.8347*** | 0.8345*** | 0.8347*** | 0.8346*** | 0.8345*** |
| | (0.1613) | (0.1614) | (0.1613) | (0.1613) | (0.1614) |
| Unemployment rate | -0.0143*** | -0.0143*** | -0.0143*** | -0.0143*** | -0.0143*** |
| | (0.0033) | (0.0033) | (0.0033) | (0.0033) | (0.0033) |
| Bank × Quarter × Year FE | Yes | Yes | Yes | Yes | Yes |
| Branch FE | Yes | Yes | Yes | Yes | Yes |
| Adj. $R^2$ | 0.944 | 0.944 | 0.944 | 0.944 | 0.944 |
| Observations | 98,951 | 98,951 | 98,951 | 98,951 | 98,951 |

Notes: This table reports estimates of equation (4). Variable definitions are reported in Table 1. The unreported control variable are income growth, population, establishments per capita, unemployment rate, bank size, capital ratio, and branches. The standard errors are two-way clustered at the bank and quarter-year levels and are reported in parentheses. *, **, and *** indicates statistical significance at the 10%, 5%, and 1% levels, respectively.

Table 6.A: Large bank placebo tests

| | 1 | 2 | 3 | 4 |
|---|---|---|---|---|
| | 100 largest banks | | 1,000 largest banks | |
| Dependent variable | Funding Rate | APY | Funding Rate | APY |
| Fintech Index | 0.0009 | 0.0010 | 0.0052 | 0.0053 |
| | (0.0034) | (0.0034) | (0.0037) | (0.0037) |
| Income growth | -0.0022** | -0.0022** | 0.0003 | 0.0003 |
| | (0.0008) | (0.0009) | (0.0011) | (0.0011) |
| Population | 0.0590 | 0.0630 | 0.0026 | 0.0026 |
| | (0.0803) | (0.0812) | (0.0026) | (0.0026) |
| Establishment per capita | -0.0321 | -0.0290 | -0.0096 | -0.0096 |
| | (0.0873) | (0.0876) | (0.0195) | (0.0197) |
| Unemployment Rate | 0.0001 | 0.0001 | -0.0007 | -0.0007 |
| | (0.0021) | (0.0022) | (0.0017) | (0.0018) |
| Bank x Quarter x Year FE | Yes | Yes | Yes | Yes |
| Branch FE | Yes | Yes | Yes | Yes |
| Adj. $R^2$ | 60,069 | 60,069 | 128,397 | 128,397 |
| Observations | 0.9665 | 0.9666 | 0.9358 | 0.9360 |

Notes: This table reports estimates of placebo experiment using equation (1) for subsample of 100 largest banks (1 & 2) and 1000 big banks (3 & 4). Variable definitions are reported in Table 1. The standard errors are two-way clustered at the bank and quarter-year levels and are reported in parentheses. *, **, and *** indicates statistical significance at the 10%, 5%, and 1% levels, respectively.

Table 7.A: Cannabis Falsification Test

| Dependent variable | 1 | 2 | 3 | 4 |
|---|---|---|---|---|
| | Funding rate | | APY | |
| Fintech Index | | 0.0297** | | 0.0296** |
| | | (0.0126) | | (0.0127) |
| Cannabis Legislation | 0.0058 | 0.0025 | 0.0060 | 0.0026 |
| | (0.0223) | (0.0217) | (0.0221) | (0.0215) |
| Income growth | -0.0065*** | -0.0060*** | -0.0065*** | -0.0060*** |
| | (0.0019) | (0.0018) | (0.0019) | (0.0018) |
| Population | 0.0625* | 0.0604 | 0.0635* | 0.0613 |
| | (0.0357) | (0.0374) | (0.0357) | (0.0373) |
| Establishment per capita | 0.8429*** | 0.7890*** | 0.8515*** | 0.7976*** |
| | (0.2771) | (0.2632) | (0.2775) | (0.2635) |
| Unemployment Rate | -0.0078 | -0.0089 | -0.0079 | -0.0090 |
| | (0.0076) | (0.0076) | (0.0076) | (0.0077) |
| Bank x Quarter x Year FE | Yes | Yes | Yes | Yes |
| Branch FE | Yes | Yes | Yes | Yes |
| Adj. $R^2$ | 0.9191 | 0.9192 | 0.9193 | 0.9195 |
| Observations | 208,171 | 208,171 | 208,171 | 208,171 |

Notes: This table reports estimates using equation (1) with additional Cannabis Legislation. Variable definitions are reported in Table 1. The standard errors are two-way clustered at the bank and quarter-year levels and are reported in parentheses. *, **, and *** indicates statistical significance at the 10%, 5%, and 1% levels, respectively.

Table 8.A: Sensitivity Checks

| | 1 | 2 | 3 | 4 | 5 | 6 | 7 | 8 |
|---|---|---|---|---|---|---|---|---|
| Dependent variable: Funding rate | | | | | | | | |
| Sample | All | 2004-2011 | All | All | All | All | All | 2011-2019 |
| Fintech index | 0.0303*** | 0.0221*** | 0.0283*** | 0.0309*** | 0.0246*** | 0.0771*** | 0.0437*** | 0.0246*** |
| | (0.0035) | (0.0033) | (0.0038) | (0.0034) | (0.0041) | (0.0104) | (0.0104) | (0.0056) |
| Fintech index ×Equity-crowdfunding regulation | -0.0024 | | | | | | | |
| | (0.0034) | | | | | | | |
| Fintech index × FolioFn | | | 0.0022 | | | | | |
| | | | (0.0033) | | | | | |
| Fintech index × VC deal per capita | | | | -0.0001*** | | | | |
| | | | | (0.0000) | | | | |
| Fintech index ×Corporate tax rate | | | | | 0.0010** | | | |
| | | | | | (0.0005) | | | |
| Fintech index ×Housing index | | | | | | -0.0002*** | | |
| | | | | | | (0.0000) | | |
| Fintech index ×Auto delinquency rate | | | | | | | 0.0019 | |
| | | | | | | | (0.0019) | |
| Fintech index × CC delinquency rate | | | | | | | -0.0043*** | |
| | | | | | | | (0.0014) | |
| Fintech index × Mortgage delinquency rate | | | | | | | 0.0002 | |
| | | | | | | | (0.0009) | |
| Fintech index ×Student loan delinquency rate | | | | | | | 0.0019** | |
| | | | | | | | (0.0008) | |
| Control Variables | Yes | Yes | Yes | Yes | Yes | Yes | Yes | Yes |
| Bank × Quarter × Year FE | Yes | Yes | Yes | Yes | Yes | Yes | Yes | Yes |
| Branch FE | Yes | Yes | Yes | Yes | Yes | Yes | Yes | Yes |
| Adj. R² | 0.929 | 0.925 | 0.929 | 0.929 | 0.929 | 0.930 | 0.929 | 0.819 |
| Observations | 208,171 | 117,998 | 208,171 | 208,171 | 208,171 | 208,171 | 208,171 | 105,103 |

Notes: This table reports estimates of equation (1). Variable definitions are reported in Table 1. The unreported control variables are income growth, population, establishments per capita, unemployment rate, bank size, capital ratio, and branches. The standard errors are two-way clustered at the bank and quarter-year levels and are reported in parentheses. *, **, and *** indicates statistical significance at the 10%, 5%, and 1% levels, respectively.

Table 9.A: Sensitivity Checks

| Sample | 1<br>All | 2<br>2004-2011 | 3<br>All | 4<br>All | 5<br>All | 6<br>All | 7<br>All | 8<br>2011-2019 |
|---|---|---|---|---|---|---|---|---|
| Fintech index | 0.0303***<br>(0.0035) | 0.0220***<br>(0.0034) | 0.0282***<br>(0.0038) | 0.0309***<br>(0.0034) | 0.0245***<br>(0.0041) | 0.0777***<br>(0.0104) | 0.0435***<br>(0.0105) | 0.0245***<br>(0.0056) |
| Fintech index ×Equity-crowdfunding regulation | -0.0024<br>(0.0034) | | | | | | | |
| Fintech index × FolioFn | | | 0.0022<br>(0.0033) | | | | | |
| Fintech index × VC deal per capita | | | | -0.0001***<br>(0.0000) | | | | |
| Fintech index ×Corporate tax rate | | | | | 0.0010**<br>(0.0005) | | | |
| Fintech index ×Housing index | | | | | | -0.0002***<br>(0.0000) | | |
| Fintech index ×Auto delinquency rate | | | | | | | 0.0019<br>(0.0020) | |
| Fintech index × CC delinquency rate | | | | | | | -0.0043***<br>(0.0014) | |
| Fintech index × Mortgage delinquency rate | | | | | | | 0.0001<br>(0.0009) | |
| Fintech index ×Student loan delinquency rate | | | | | | | 0.0019**<br>(0.0008) | |
| Control Variables | Yes | Yes | Yes | Yes | Yes | Yes | Yes | Yes |
| Bank × Quarter × Year FE | Yes | Yes | Yes | Yes | Yes | Yes | Yes | Yes |
| Branch FE | Yes | Yes | Yes | Yes | Yes | Yes | Yes | Yes |
| Adj. $R^2$ | 0.929 | 0.925 | 0.929 | 0.929 | 0.929 | 0.930 | 0.929 | 0.819 |
| Observations | 208,171 | 117,998 | 208,171 | 208,171 | 208,171 | 208,171 | 208,171 | 105,103 |

Notes: This table reports estimates of equation (1). Variable definitions are reported in Table 1. The unreported control variable are income growth, population, establishments per capita, unemployment rate, bank size, capital ratio, and branches. The standard errors are two-way clustered at the bank and quarter-year levels and are reported in parentheses. *, **, and *** indicates statistical significance at the 10%, 5%, and 1% levels, respectively.

Table 10.A: Consumer Loan Interest Ceilings

| | 1 | 2 | 3 | 4 |
|---|---|---|---|---|
| Deposit measure | Funding rate | | APY | |
| Sample | Interest ceiling | No interest ceiling | Interest ceiling | No interest ceiling |
| Fintech Index | 0.0274*** | 0.0385** | 0.0276*** | 0.0385** |
| | (0.0047) | (0.0145) | (0.0047) | (0.0145) |
| Income growth | -0.0116*** | -0.0039 | -0.0117*** | -0.0039 |
| | (0.0014) | (0.0026) | (0.0014) | (0.0026) |
| Population | 0.2891*** | 0.0300 | 0.2883*** | 0.0300 |
| | (0.0547) | (0.0303) | (0.0547) | (0.0303) |
| Establishment per capita | 0.9966*** | 1.0268*** | 0.9969*** | 1.0268*** |
| | (0.1296) | (0.3359) | (0.1298) | (0.3359) |
| Unemployment Rate | 0.0101*** | -0.0157* | 0.0103*** | -0.0157* |
| | (0.0032) | (0.0085) | (0.0032) | (0.0085) |
| Bank x Quarter x Year FE | Yes | Yes | Yes | Yes |
| Branch FE | Yes | Yes | Yes | Yes |
| Adj. R² | 0.9194 | 0.9197 | 0.9197 | 0.9197 |
| Observations | 78,834 | 125,763 | 78,834 | 125,763 |

Notes: This table reports estimates of equation (1) for sample spits into states with and without interest ceiling. Variable definitions are reported in Table 1. The control variables are income growth, population, establishments per capita, unemployment rate, bank size, capital ratio, and branches. The standard errors are two-way clustered at the bank and quarter-year levels and are reported in parentheses. *, **, and *** indicates statistical significance at the 10%, 5%, and 1% levels, respectively.